\documentclass[useAMS,usenatbib,usegraphicx,galley]{mn2e}

\usepackage{subfigure}
\usepackage{amssymb}
\usepackage{balance}
\usepackage{ulem} 
\newcommand{\vspc}{\vspace{10pt}}
\newcommand{\Msun}{M_{\odot}}

\usepackage{fancyhdr}
\begin{document}
\normalem 

\date{20 March 2010}


\title{3D-Matched-Filter Galaxy Cluster Finder I: \\ Selection Functions and CFHTLS Deep Clusters}
\author[Milkeraitis et al.]
  {M. Milkeraitis$^1$\thanks{martham@phas.ubc.ca}, L. Van Waerbeke$^1$, C. Heymans$^2$, H. Hildebrandt$^3$, \newauthor J. P. Dietrich$^{4,5,6}$, T. Erben$^7$\\
  $^1$University of British Columbia, Department of Physics and Astronomy, 6224 Agricultural Rd, Vancouver BC, V6T 1Z1, Canada \\
  $^2$ The Scottish Universities Physics Alliance, Institute for Astronomy, University of Edinburgh, Blackford Hill, Edinburgh EH9 3HJ, UK\\ 
  $^3$ Leiden Observatory, Leiden University, Niels Bohrweg 2, 2333CA Leiden, The Netherlands \\ 
  $^4$ ESO, Karl-Schwarzschild-Str. 2, 85748 Garching b. M\"unchen, Germany\\
  $^5$ Physics Department, University of Michigan, 450 Church St, Ann Arbor, MI 48109-1040, USA \\
  $^6$ Michigan Center for Theoretical Physics, 450 Church St, Ann Arbor, MI 48109-1040, USA \\
  $^7$ Argelander-Institut f\"ur Astronomie, Auf dem H\"ugel 71, 53121 Bonn, Germany\\
}
\maketitle

\begin{abstract} 
We present an optimised galaxy cluster finder, 3D-Matched-Filter (3D-MF), that utilises galaxy cluster radial profiles, luminosity functions and redshift information to detect galaxy clusters in optical surveys.  This method is an improvement over other matched-filter methods, most notably through implementing redshift slicing of the data to significantly reduce line-of-sight projections and related false positives.  We apply our method to the Canada-France-Hawaii Telescope Legacy Survey Deep fields, finding $\sim 170$ galaxy clusters per square degree in the $0.2 \le z \le 1.0$ redshift range.  Future surveys such as LSST and JDEM can exploit 3D-MF's automated methodology to produce complete and reliable galaxy cluster catalogues.  We determine the reliability and accuracy of the statistical approach of our method through a thorough analysis of mock data from the Millennium Simulation.  We detect clusters with 100\% completeness for $M_{200} \ge 3.0\times 10^{14}\Msun$, 88\% completeness for $M_{200} \ge 1.0\times 10^{14}\Msun$, and 72\% completeness well into the $10^{13}\Msun$ cluster mass range.  We show a 36\% multiple detection rate for cluster masses $\ge 1.5\times 10^{13}\Msun$ and a 16\% false detection rate for galaxy clusters $\gtrsim 5\times 10^{13}\Msun$, reporting that for clusters with masses $\lesssim 5\times 10^{13}\Msun$ false detections may increase up to $\sim24\%$.  Utilising these selection functions we conclude that our galaxy cluster catalogue is the most complete CFHTLS Deep cluster catalogue to date.
\end{abstract}


\section{Introduction}\label{intro}
Clusters of galaxies pinpoint the densest regions in the Universe.  Housed in deep gravitational wells, clusters act as laboratories to study the influence of extreme environments on galaxy formation and evolution \citep{Dressler80}.  The most massive clusters are also natural gravitational telescopes, lensing and magnifying light from the most distant of galaxies \citep{Stark07}.  As clusters trace the high mass tail of the matter distribution, a complete sample can provide a very sensitive probe of cosmology \citep{BW03}.  Directly probing the growth of structure in the early Universe, cluster cosmology is fast becoming an important component of future dark energy surveys.  As the future of optical astronomy ushers in large datasets of wide, deep sky coverage, it becomes necessary to develop automated algorithms that methodically and accurately search these datasets for galaxy clusters.  Depending on scientific goals, it is desirable to have as complete a galaxy cluster sample as possible, over a range of redshifts; any intrinsic limitations of these search algorithms and the resultant biases introduced into generated galaxy cluster catalogues must be understood.  The question of how to methodically select and quantify the completeness of a cluster sample is the subject of this paper.\\
\\
Many different approaches to searching for galaxy clusters in astronomical data exist: current optical astronomy methods focus on finding clusters as overdensities via friends-of-friends algorithms such as \cite{tornadoli}, density maps as in \cite{adami}, or Voronoi tesselation methods as in \cite{voronoilopes}, and \cite{twotecx} for example.  Other search methods look for large, red, elliptical cluster galaxy populations, and are referred to as red sequence techniques (\cite{gladdersyee2000}, \cite{cohn}, \cite{kodama}, \cite{tinglu}, and \cite{k2ref} for example), perhaps also including the existence of bright central galaxies into the search algorithms (such as the maxBCG method of \cite{koester}, and \cite{hilbert}).  Alternatively, other methods search for clusters based on characteristic galaxy cluster luminosity and radial profiles (\cite{postman}, \cite{olsen1stEISpaper}, \cite{kepner}, \cite{whitekochanek}, \cite{KW03}, \cite{gilbank}, \cite{joergd1}, \cite{grove2ndcfhtls}, \cite{menanteau09}).  Each of these methods makes assumptions about general cluster properties and resultantly, derived galaxy cluster catalogues will primarily include galaxy clusters that reflect those assumptions.  Ideally a method that minimises these biases and produces an understood and complete galaxy cluster catalogue will be best suited for statistical science of galaxy clusters and investigations into cosmology. \\
\\
With this in mind, we present an optimised luminosity function and radial profile based optical galaxy cluster finding algorithm that we call 3D-Matched-Filter (3D-MF).  We present thorough tests of 3D-MF on simulations, and use the ascertained galaxy cluster selection functions to produce a galaxy cluster catalogue for the Canada-France-Hawaii Telescope Legacy Survey Deep fields.  To this effect, Section \ref{MS} of this paper will discuss the Millennium Simulation catalogues used, Section \ref{MF} will discuss matched-filter methodology, as well as 3D-MF itself, and Section \ref{3DMFonMSnozerr} will outline the selection functions of 3D-MF from the Millennium Simulation data ({\emph{without}} and {\emph{with}} photometric redshift error).  Subsequently, 3D-MF will be run on the Canada-France-Hawaii Telescope Legacy Survey Deep Dataset and the resultant galaxy cluster catalogues will be discussed in Section \ref{3DMFonCFHTLS}.  \\
\\
A $\Lambda$CDM cosmology has been assumed throughout this work: $H_0=73$ km$/$s$/$Mpc, $\Omega_M=0.25$, $\Omega_{\Lambda}=0.75$ (which is also consistent with the Millennium Simulation used herein). 

\section{Millennium Simulation}\label{MS}
\subsection{Catalogues}
\noindent The simulation catalogues used throughout this work are six pencil-beam mock catalogues from \cite{MSkitzbichler2007} (henceforth KW07).  These catalogues are lightcones created from the semi-analytic galaxy catalogue from \cite{MSdelucia2007}, which in turn is derived from the Millennium Simulation of the concordance $\Lambda$CDM cosmogony (\cite{MSspringel2005}; henceforth S05). \\
\\
The Millennium Simulation well-reproduces galaxy clustering as a function of luminosity as shown in Figure 5 of S05.  Furthermore, semi-analytic modelling applied to the original Millennium Simulation catalogues, in the \cite{MSdelucia2007} and KW07 catalogues, results in luminosity functions in similarly good agreement with observations.  Consequently, these catalogues are well suited to test a cluster finding algorithm based on luminosity, such as the 3D-MF galaxy cluster finding algorithm presented in this work.  Conversely, there are difficulties in reproducing observational colours in simulation work (again, see Figure 5 from S05 and a discussion in \cite{MSdelucia2007}).  This fact hinders the analysis of colour-based cluster finding algorithms, such as red sequence based techniques, on the Millennium Simulation.  \cite{cohn} tested their red sequence algorithm to examine changes with redshift using the Millennium Simulation.  The authors found the simulation colours did not match observations; they instead determined their red sequence by matching the simulations.  In another recent example, \cite{hilbert} used the maxBCG method from \cite{koester} (based on the red sequence technique) on the Millennium Simulation and found that the simulation did not reproduce the colours of passively evolving galaxies to the degree required for a direct application of colour-redshift relations derived from observations.  Resultantly, they used altered colour-redshift relations, noting that if they had not adjusted the colour-redshift selection relations by hand, running maxBCG on the Millennium Simulation dataset and subsequent modelling would have found almost no clusters with $z>0.25$.  The 3D-MF technique does {\emph{not}} use Millennium Simulation colour information, and relies instead on the luminosity function of galaxy clusters; therefore, evaluating on the Millennium Simulation provides a robust test of the application of this method to real data.\\ 
\\
The KW07 mock catalogues used herein consist of six deep fields of $1.4 \times 1.4$ square degrees and mark the first time these lightcones have been used in cluster finding work.  3D-MF uses redshift information: this is not obtained from Millennium Simulation colours.  Magnitude information is used from KW07 but only for the purpose of calculating luminosity functions as described.  The SDSS magnitudes were converted to MegaCam magnitudes (according to Equation \ref{magchange}\footnote{The Canadian Astronomy Data Centre, Herzberg Institute of Astrophysics, MegaCam ugriz filter set as per:\\ http://www2.cadc-ccda.hia-iha.nrc-cnrc.gc.ca/megapipe/ \\docs/filters.html}) in anticipation of the Canada-France-Hawaii Telescope data that 3D-MF will first have the opportunity to examine in Section \ref{3DMFonCFHTLS}.
\begin{equation}\label{magchange}
i_{\mathrm{MegaCam}} = i_{\mathrm{SDSS}} - 0.085 \ (r_{\mathrm{SDSS}} - i_{\mathrm{SDSS}})
\end{equation}
Hereafter we refer to $i_{\mathrm{MegaCam}}$ as $i'-$band.

\subsection{Millennium Simulation Galaxy Clusters}\label{MSgalcls}
\begin{figure}
  \includegraphics[width=8cm]{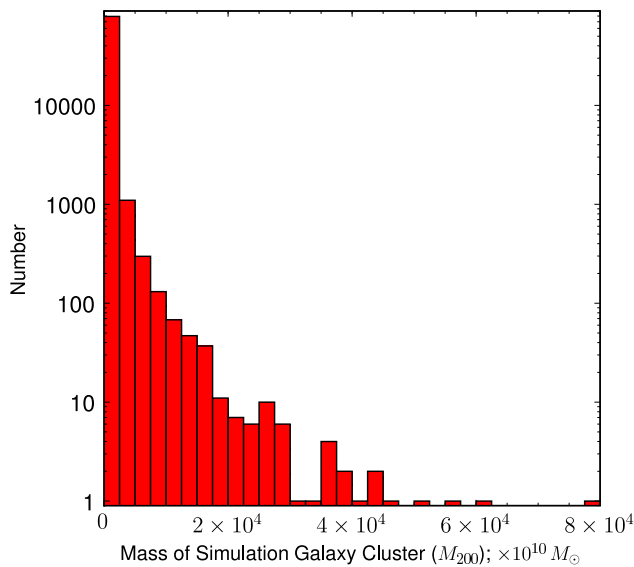}
  \caption{Mass distribution of Millennium Simulation KW07 galaxy clusters (following the cluster definition given in Table \ref{MS_CLdef}, and showing the cluster redshift range of $0.2 \le z \le 1.0$ as explained in the text).}
      \label{MSallclustersMass}
\end{figure}

\begin{figure}
  \includegraphics[width=8cm]{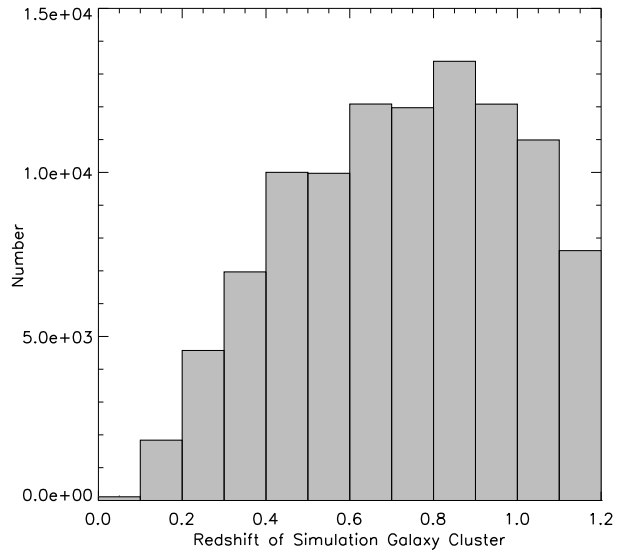}
  \caption{Redshift distribution of Millennium Simulation KW07 galaxy clusters (following the cluster definition given in Table \ref{MS_CLdef}, and noting the catalogue lightcones extend much further than a redshift of 1.2: higher ranges are not shown here).}
      \label{MSallclustersnz}
\end{figure}

A common definition does not exist in the literature with which to define a cluster, other than the agreed upon fact that a cluster consists of multiple galaxies gravitationally bound to one another orbiting a common centre of mass.  Smaller numbers of galaxies bound together are more traditionally known as {\emph{groups}}, whereas larger galaxy associations are typically called {\emph{clusters}}; it is the border between groups and clusters that is not always strictly defined.  For the purposes of this work, a cluster will be defined in the Millennium Simulation as a collection of galaxies that belong to the same parent halo and have $\ge 5$ members (where membership is defined by examining the Millennium Simulation catalogues to $z<1.2$); traits of {\emph{clusters}} are listed in Table \ref{MS_CLdef}.  Note that $M_{200}$ is the mass enclosed within radius $r_{200}$: the radius within which the mean density is 200 times the critical density at that redshift.  Based on this galaxy cluster definition, the mass distribution of KW07 galaxy clusters is shown in Figure \ref{MSallclustersMass}.  The redshift distribution of these {\emph{known}} clusters is presented in Figure \ref{MSallclustersnz}.  \\
\begin{table}
  \caption{Parameters used to compile a list of known galaxy clusters from the KW07 mock lightcones of the Millennium Simulation. \label{MS_CLdef}}
   \begin{tabular*}{0.44\textwidth}{@{\extracolsep{\fill}}l}
         \hline
	 \\ [-2ex]
	 \multicolumn{1}{l}{\bf Millennium Simulation Galaxy Cluster Traits} \\ [1ex] \hline
	 \\ [-2ex]
         \ \textbullet \ \ Cluster members have the same friends-of-friends \\
identification number (as described in the simulations), \\ thus avoiding dark matter substructures \ \\ [0.5ex] \hline
	 \\[-2ex]
         \ \textbullet \ \ Cluster members are flagged in the simulations \\ with the $M_{200}$ of their parent halo \\ [0.5ex] \hline 
	 \\[-2ex]
         \ \textbullet \ \ Clusters have $\ge 5$ galaxy members \\ [0.5ex] \hline \\
      \end{tabular*}
\end{table} 


\noindent The centre for each known Millennium Simulation cluster is defined in this work as the average position of the brightest 10 (or average of all members when the number of galaxies per cluster is $< 10$) $i'-$band objects per cluster.  The position of the centre is only used when matching the clusters found by 3D-MF to those known in the Millennium Simulation to determine which clusters were found, and a tolerance range is used (described in Section \ref{3DMFonMSnozerr}); it is not crucial to immediately determine a more exact position of the cluster centre and it will be examined further in Section \ref{centroidsec}.

\section{The 3D-MF Algorithm}\label{MF}
\subsection{The Foundations}\label{oldfoundations}
The 3D-MF galaxy cluster finding algorithm bases its search on the luminosity and radial profile of a galaxy cluster, appropriately sized for the redshift of the cluster.  Prior to implementing our changes (described in Section \ref{newimp}), this method was based on \citep[][hereafter P96]{postman}, and the reader is directed there for a more in-depth discussion of the background of this technique.  The method can use any sensible choice of luminosity or radial profile to model a galaxy cluster: 3D-MF follows P96 and currently uses a modified Schechter luminosity function (Equation \ref{eqn:schec}) and truncated Hubble radial profile (Equation \ref{eqn:hub}) to describe a fiducial galaxy cluster: 
\vspc
\begin{equation}\label{eqn:schec}
  \Phi(M)=0.4 \ \ln10 \ \Phi^* 10^{0.4(\alpha + 1)(M^*-M)} \exp \left [-10^{0.4(M^*-M)} \right ]
\end{equation}

\indent where $\Phi$ is the galaxy luminosity function, $\Phi^*$ sets the overall normalisation of galaxy density, $M$ is absolute magnitude, and $\alpha$ is the slope of the faint end of the luminosity function.  $M^*$ is a characteristic absolute magnitude: $\Phi$ drops with increasing luminosity but at magnitudes brighter than $M^*$ the exponentially decreasing slope of the luminosity function cuts off dramatically.  The integral of the Schechter luminosity function diverges for $\alpha < -1$ at magnitudes fainter than $M^*$; the multiplicative term, exp$[-10^{0.4(M^{*}-M)}]$, that has been added to modify the Schechter equation, is a dimensionless power-law cutoff (weighted by the flux of the galaxy) that allows the function to remain integrable. \\
\\
The radial profile of the cluster is modelled as a projected truncated Hubble radial profile;
\vspc
  \begin{equation}\label{eqn:hub}
    P \left ( \frac{r}{r_c} \right )=\frac{1}{\sqrt{1+ \left ( \frac{r}{r_c} \right ) ^2}} - \frac{1}{\sqrt{1+ \left ( \frac{r_{co}}{r_c} \right ) ^2}}
  \end{equation}

\indent where $r$ is the distance of a galaxy from a cluster centre, $r_c$ is the cluster core radius, and $r_{co}$ is the cutoff radius.  $r_{co}$ is arbitrary and must be chosen such that it extends well beyond the cluster's core radius, importantly allowing non-circularly symmetric clusters to still be detected by this technique. \\
\\
Once the fiducial cluster has been established by choice of luminosity and radial profile, the data is searched in one wavelength band for areas that maximally and simultaneously match both profiles, thereby {\emph{filtering}} galaxy clusters out of the data (and thus the terminology 'matched-filter').  In previous applications of matched-filter methodologies, the data have not been binned per redshift as photometric redshift information has only recently become commonplace.  In the absence of photometric redshift information in the past, the profiles that were being matched were themselves re-determined with an assumed cluster size and $M^*$ value at a series of trial redshifts. \\
\\
The likelihood, $\mathcal{L}$, that galaxies within $r_{co}$ match the luminosity and radial profile of a fiducial cluster at a particular assumed redshift is calculated according to Equation \ref{eqn:likeli}:
\vspc
\begin{equation}\label{eqn:likeli}
  \ln \ \mathcal{L} \propto \int P \left ( \frac{r}{r_c} \right ) \frac{\Phi(m-m^*)}{b(m)} D(r,m) \ \mathrm{d}^2r\ \mathrm{d}m
\end{equation}

\indent where $m$ is the apparent magnitude of a galaxy, $m^*$ is the apparent magnitude corresponding to the characteristic luminosity of cluster galaxies, which incidentally includes a redshift dependent {\emph{k}}-correction, $b(m)$ is the background galaxy count, and $D(r,m)$ is the total number of galaxies at a given magnitude and distance from a cluster centre.  \\
\\
This likelihood is calculated for every possible cluster centre in an input catalogue and can be viewed as an array, or image map; these maps will be referred to as likelihood maps.  The likelihood maps are searched for peaks, which are galaxy cluster detections.  \\
\\
The significance of a peak, or galaxy cluster detection, is measured according to Equation \ref{eqn:sigma}:
\begin{equation}\label{eqn:sigma}
  \sigma = (S_p - S_{bg}) / \sigma_{bg}
\end{equation}

\indent where $S_p$ is the peak signal, and $S_{bg}$ is the background signal calculated by binning the distribution of all likelihood values in a likelihood map, and taking the mode of this Gaussian distribution.  The background dispersion, $\sigma_{bg}$, is $0.741 \times (Q_3 - Q_1)$ where $Q_1$ and $Q_3$ are the first and third quartiles of the ranked likelihood value distribution.  Cluster detections can be filtered based on their significance, ensuring detections are well above the background noise, and in this way a galaxy cluster list can be generated.

\subsection{2D False Detection Rates}\label{oldhighfalse}
False detections are defined as galaxy clusters found by an algorithm that do not exist in reality: an important parameter, among others, by which to define the purity of an algorithm.  Older versions of matched-filter techniques have uncomfortably high false detection rates, a feature largely attributable to line-of-sight projections.\\
\\
P96 test their matched-filter algorithm on a self-created simulation (mimicking the Palomar Distant Cluster Survey) and find a 65\% false detection rate of galaxy clusters with peak signal thresholds $\ge 3 \sigma$.  P96 reduce this false detection rate to 31\% by adding the criteria that clusters must be found by their algorithm twice, looking separately in 2 colour bands.  The ESO Imaging Survey had some success in finding clusters using their implementation of this same algorithm, but it was initially developed with the purpose of preparing a list of candidate clusters for follow-up observations and not for producing a well-defined sample for statistical analysis (\cite{olsen1stEISpaper}, \cite{olsen2ndEISpaper}; O1\&2).  These ESO Imaging Survey cluster papers, and subsequent work with their algorithm (\cite{joergd1}: D07; and \cite{olsen1stcfhtls}, \cite{grove2ndcfhtls}: O3\&4), have not tested their version of the matched-filter algorithm on simulations in terms of its ability to recover known clusters, other than to very generally quantify the ability of their finder to avoid noise detections. Furthermore, these papers aim for the same galaxy cluster density as the Palomar Distant Cluster Survey in P96 (i.e. O3 quotes $\sim 52.1$ clusters per square degree), and use the same method, implying a similar false detection rate to that quoted above (O3 claims $16.9\pm 5.4$ false detections per square degree, or $33\%$).\\
\\
\cite{kepner} implement a matched-filter algorithm based on P96 as well, but do not examine cluster recovery numbers.  Instead, they focus on the accuracy of their algorithm to detect a given cluster at its proper redshift, or the accuracy of determining the proper richness of a given cluster.  White and Kochanek implement a matched-filter algorithm based on Kepner's version \citep[][hereafter WK02]{whitekochanek} and test on self-created simulations to examine their cluster finding abilities; however, they find their simulations contain too few galaxies compared to real observational surveys: in fact, 60\% too few galaxies for an $R$-band magnitude $<24$.  This limitation affects complications caused by line-of-sight projections, as well as luminosity functions integral to the matched-filter methodology, and thus efficiency rates of their algorithm compared to real data are reported as conservative estimates.  \cite{kepner} and WK02 use different likelihood functions from Equation \ref{eqn:likeli} in this work, as well as different peak detection methods, and the reader is directed to those papers for more details.  WK02 make no adjustments to improve cluster detections at the edges of their data.  These edge effects become important when using cluster finders on real data: masking bad pixels or bright stars is necessary in reality and clusters on the borders of the masked regions, if not properly weighted or accounted for in terms of the fraction available to the search algorithm, will likely be missed.  In terms of false detection reports, WK02 show how their false detection rates change with redshift: for their deepest self-created simulation of 1.5 x 1.5 square degrees out to redshift 1.0, with a limiting R band magnitude of 24, they have false positive rates of $>40\%$ (averaging $\approx60\%$ for all redshift bins) for clusters $\ge 2 \times 10^{14}\Msun$.\\
\\
Proponents of other cluster finding techniques, such as red sequence based methods, often comment upon older matched-filter methods' high false detection rates as a reason to avoid this technique (\cite{gladdersyee2000} or \cite{gilbank} for example).  However, the false detection rates quoted above are largely caused by line-of-sight projections in the data; removing this contamination will lead to much improved false detection rates and push down the galaxy cluster mass limit to which reliable detections can be made.  Arguably, red sequence based methods have their own line-of-sight projection issues.  \cite{gladdersyee2000} for example introduced their red sequence based cluster finder by testing it on known CNOC2 clusters (i.e. not simulations, but real data), and found excellent false positive rates of $<5\%$ for clusters at redshifts $<0.5$.  However, as WK02 also show, false positives increase with redshift.  The \cite{gladdersyee2000} CNOC2 sample testing provides limitations in the higher redshift ranges: notably a 36\% false detection rate in the redshift range $0.5<z<0.6$ (where authors suggest spectroscopic data was incomplete and did not deduce anything further).  Furthermore, \cite{koester}, using their maxBCG method (based on the red sequence method), find 16\% of found galaxy clusters with shallow redshifts ($0.1<z<0.3$) are in reality line-of-sight projections (as tested on the observational dataset of the SDSS).  As all optical cluster finders have false positive rates that will increase with redshift, it becomes a goal of this work to minimise these false detections as much as possible.\\
\\
Intrinsic biases will exist to some degree in all methods.  For example, red sequence based methods have their own intrinsic biases about galaxy clusters, notably assuming all galaxy clusters have large enough red sequence populations to be detected based on red sequence alone.  Complimentary galaxy cluster finding techniques with different biases are therefore necessary, first to determine whether clusters can be found that are missed by other techniques, and second to be able to ultimately construct a galaxy cluster list from observational data that is as complete as possible, and whose biases are well-understood (via testing on simulations). 

\subsection{3D-MF's Advancements}\label{newimp} 
Using the foundations of the method discussed above in Section \ref{oldfoundations}, 3D-MF advances the matched-filter technique by incorporating photometric redshift information, and adjusts additional elements as described below, to update and improve galaxy cluster finding.  With the requirement that all next-generation surveys for cosmology come with photometric redshifts, it is natural for 3D-MF to extend the matched-filter technique into this third dimension.

\subsubsection{Optimal Use of Photometric Redshifts}\label{zslicingexplained}
3D-MF cuts the input data by redshift, initially into non-overlapping redshift slices.  Each slice is then matched against appropriate luminosity and radial filters scaled for that particular redshift slice.  The resultant likelihood maps (described in Section \ref{oldfoundations} with Equation \ref{eqn:likeli}) are run individually through a new 3D-MF peak detection pipeline to identify galaxy clusters.  Upon completion of a range of sequential redshift slices the entire process is repeated with a shift in the slices (thus overlapping the redshift slicing with the initial run) in order to account for clusters or structures that exist on borders of the original slicing.  For example, an input catalogue could be sliced from redshift 0.00 to 1.20 (with a bin width of 0.2, but any desired bin width could be chosen) and these would be considered the {\emph{original}} slices.  After running 3D-MF on these original slices, the process is repeated by slicing the whole input catalogue differently, from redshift 0.10 to 1.30 (with the same bin width for example), and these would be considered the {\emph{shifted}} slices.  The resulting galaxy cluster lists from a complete run of original slices and a complete run of shifted slices are then merged (the details of which are presented in Section \ref{3DMFonMSnozerr}).  Clusters that are multiply detected can be tracked, and false detections in either the original or shifted slices can be lowered by associating them properly with clusters from the shifted or original slices respectively.  The aim of 3D-MF is to optimise cluster detections through redshift slicing, thus eliminating line-of-sight projections, but also to take into account the fact that each cluster spans a range of space and subsequently redshift slicing has to be applied carefully to not cut these clusters and count them multiple times, or conversely to miss them entirely.  In optimising 3D-MF a variety of slicing styles were examined (various slice widths, non-uniform slice widths, differing overlapping slice amounts, etc), and the slicing method that produced the best results (as seen in Section \ref{finefine}) used redshift slice widths of 0.2, with shifted slices overlapping original slices by a redshift of 0.1.

\subsubsection{Masks}\label{masks}
3D-MF accounts for masked regions of data (obscured by bad pixels, star diffraction spikes, or data regions of low signal-to-noise for example) by rejecting potential cluster candidates that are masked above a set percent (given as an input threshold, chosen here to be 50\%).  In cases where an important fraction of a cluster's luminosity function is missing due to a mask, it simply cannot be rebuilt, and missing optical information such as this is a limitation of optical galaxy cluster finding algorithms, including matched-filter methods. \\
\\
Partially masked clusters are weighted by 3D-MF according to the fraction of the cluster that is masked (Equation \ref{eqn:mask});  
\begin{equation}\label{eqn:mask}
  \mathcal{L}_{\mathrm{Mask \ weighted}} = \mathcal{L} \ (1+ \% \mathrm{masked} / 100 )
\end{equation}
\indent \hspace{5pt} where the likelihood, $\mathcal{L}$ (from Equation \ref{eqn:likeli}), is up-weighted by the percentage of the redshift-scaled radial filter that is masked, giving $\mathcal{L}_{\mathrm{Mask \ weighted}}$. \\
\\
The mask-weighted likelihood maps, $\mathcal{L}_{\mathrm{Mask \ weighted}}$, increase the likelihood that a partially masked cluster matches the fiducial cluster at a given redshift, as its luminosity and radial profiles are up-weighted to compensate for the information missing due to the presence of the mask.  Importantly, mask-weighting also improves detections of clusters near the edges of an image as clusters previously missed in these regions were found after the addition of this criteria. 

\subsubsection{Background Galaxy Counts}\label{bckgndcounts}
The requirement for an accurate measure of the background galaxy counts ($b(m)$ in Equation \ref{eqn:likeli}) is an issue in P96 and other implementations of matched-filter such as O1-4 discussed in Section \ref{oldhighfalse}.  These older implementations of matched-filter methodology assume {\emph{a priori}} that $b(m)$ can be modelled by a power law in $m$, or they determine $b(m)$ by creating self-simulated backgrounds \citep[following the prescription of][]{sonpeebles}.  3D-MF determines the background counts by binning the magnitude distribution of the input data and subsequently interpolates within this measured distribution to determine $b(m)$ for each galaxy (i.e. pulling out an interpolated background number count for a particular galaxy magnitude), including it as a data-driven $b(m)$ in the integral over cluster galaxies in Equation \ref{eqn:likeli}.  We find that this accurately models the likelihood of a cluster signal against a true data-derived background and increases a cluster's likelihood of detection with the advantage that our method does not rely on biased modelling of $b(m)$.

\subsubsection{Peak Detection}\label{peakdetn}
Previous matched-filter methods (implementations of P96 such as D07 and O1-4) detect peaks in the likelihood maps using SE{\small XTRACTOR}; while SE{\small XTRACTOR} is optimised for the detection of isolated peaks, it tends to blend things.  Although this is excellent for object detection, the likelihood maps created with a matched-filter algorithm are more akin to continuous maps and thus need an appropriate peak detection method. Furthermore, SE{\small XTRACTOR} is parameter dependent; we desire a robust method that withstands reasonable changes in parameter space.  3D-MF detects all peaks above a user-input threshold (such as $\ge 3.5\sigma$) and thereby generates a complete list of possible cluster candidates. 

\subsubsection{Assigned Cluster Redshift}\label{assnz}
Other versions of the matched-filter technique assign a redshift to a galaxy cluster where it is maximally detected (i.e. the redshift at which the fiducial cluster is maximally matched) but in practise this often assigns an incorrect redshift to the cluster because all data is considered at all redshifts.  Several other factors also contribute to this incorrect redshift assignment, among them the assumed background counts discussed above, the {\emph{k}}-correction (and the effect on this of the assumed evolution of various galaxy types), and the fact that the cluster signals become over-corrected via the normalisation of the luminosity filter (as in D07 and O1-4).  These methods apply a cluster signal correction factor to counteract some of these points, but resulting redshift assignments are still, on average, incorrect and thus are not robust (nor do they claim to be; see O1\&2).  \\
\\
A plethora of papers followed up on the clusters found in O1\&2 with the goal of obtaining spectroscopic redshifts of the matched-filter found cluster members (\cite{ramella00}, \cite{olsenspec02}, \cite{olsenspec02b}, \cite{olsenspec03}, \cite{olsenspec05}a, \cite{olsenspec05b}b, \cite{olsenspec08}).  Not all clusters were conclusively found in the spectroscopic analysis; of the ones that were, the spectroscopic redshifts for these systems were compared with their matched-filter found redshifts.  Unfortunately, number counts are too low to ascertain anything statistically (i.e. the sample size available to compare spectroscopic and photometric redshifts was often only a few galaxy clusters per paper).  \\
\\
Also investigating redshift comparisons, galaxy clusters in D07 have redshifts derived from their matched-filter implementation and a sample of these are compared to known spectroscopic redshifts from the literature.  These authors find their matched-filter has a bias toward lower redshifts and conclude they are underestimating the true redshift of galaxy clusters; a discrepancy that increases with higher redshifts.\\
\\
As 3D-MF requires photometric redshift information for all of the objects in a catalogue and slices the catalogue according to redshift, once each galaxy cluster is found, its associated redshift slice is automatically, immediately and more accurately known.  For example, for a redshift bin width of 0.2, with overlapping bins shifted by a redshift of 0.1 from the original bins, consider a cluster that is first detected in an original bin: the cluster is assigned the redshift of the centre of the bin and this cluster centre is thereby known to within $\pm 0.1$ in redshift (i.e. the width of the bin).  After repeating the cluster finding process with shifted slices (as described above in Section \ref{zslicingexplained}), this cluster may also be detected in an overlapping shifted slice, localising the redshift of its centre even further (to roughly $\pm 0.05$ in redshift: the width of the overlap region of the original and shifted slices).  This of course assumes that photometric redshifts are without error and unbiased.  Knowing a cluster's galaxy members (and their redshifts) leads to an even more localised determination of a cluster's redshift; this will be discussed in relation to 3D-MF in Section \ref{proxy}.

\subsubsection{Parameter Optimisation}
3D-MF has a variety of parameters that can be optimised for maximal galaxy cluster detection.  Using Millennium Simulation data to test 3D-MF has proven irreplaceable in its contribution to improving 3D-MF, in particular determining optimal parameter settings to apply to real data, assuming a similar cosmology.  A discussion of these parameter settings follows in Section \ref{3DMFonMSnozerr}. 

\section{3D-MF on Millennium Simulation Mock Lightcones}\label{3DMFonMSnozerr}
In this section we present the application of the 3D-MF algorithm to the Millennium Simulation KW07 mock catalogues.  We first assume exact redshifts are known for all galaxies, and then investigate the impact of adding photometric redshift errors to the data, on 3D-MF's selection functions.  When matching the 3D-MF found clusters to known mock clusters, we will use the known clusters above a mass of $1.5\times 10^{13}\Msun$, and focus on the redshift range $0.2 \le z \le 1.0$ for reasons that will be explained.

\subsection{3D-MF Parameters}\label{params}
The Millennium Simulation mock lightcone catalogues were cut at an $i'-$band magnitude of 25.5 to simulate a magnitude-limited survey.  The catalogues were then cut into optimal redshift slices (see Section \ref{finefine}): redshift slice widths of 0.2 and shifted slices shifted from original slices by 0.1 in redshift (i.e. original and shifted slices overlap by 0.1 in redshift).  The mock catalogues were then analysed with 3D-MF, as described in Section \ref{MF}, using the parameters listed in table \ref{3dmfparams}.\\ 

\begin{table}
  \caption{3D-MF basic run-time parameters (consult the text for descriptions of each parameter). \label{3dmfparams} }
  \begin{tabular*}{0.45\textwidth}{@{\extracolsep{\fill}}rl}
    \hline 
    \\ [-2ex]
    \multicolumn{2}{l}{\bf 3D-MF Input Parameters} \\ [1ex] \hline
    \\ [-2ex]
    \multicolumn{2}{l}{Schechter Function (Equation \ref{eqn:schec})} \\ [0.3ex] \hline
    \\[-2ex]
    $M^*_{i'-\mathrm{band}}$ & -20.69 \\ [0.8ex] 
    $\alpha$ & -1.11 \\ [1ex] \hline
    \\[-2ex]
    \multicolumn{2}{l}{Hubble Profile (Equation \ref{eqn:hub})} \\ [0.3ex] \hline
    \\[-2ex]
    $r_c$ & 0.1Mpc \\ [0.8ex] 
    $r_{co}$ & 1Mpc \\ [1.5ex] \hline
    \\[-2ex]
    \multicolumn{2}{l}{Likelihood Map (Equation \ref{eqn:likeli})} \\ [0.3ex] \hline
    \\[-2ex]
    Scale & $0.124\frac{'}{pix}$ \\ [1.5ex] \hline
    \\[-2ex]
    \multicolumn{2}{l}{Peak Detection} \\ [0.3ex] \hline
    \\[-2ex]
    Mask Weighting & $\mathcal{L} \ (1+ \% \mathrm{masked} / 100 $) \\ [1ex]  
    Cluster Rejection & if $\ge 50\%$ masked, reject \\ [1ex] 
    Minimum Significance &  $\ge 3.5 \sigma$ \\
    (Equation \ref{eqn:sigma}) for Real & \\
    Detection & \\ [0.5ex] \hline
    \\[-2ex]
    \multicolumn{2}{l}{Multiple Detections} \\ [0.3ex] \hline	 
    \\[-2ex]
    2D Grouping (RA,Dec) & 1.5Mpc diameter \\
    in Same Redshift Slice & \\ [1ex] 
    Grouping Between & $z_{\mathrm{3D-MF}}\pm 0.1$ \\
    Redshift Slices & \\ \hline
\\
  \end{tabular*}
\end{table} 
 
\noindent $M^*$ and $\alpha$ (Equation \ref{eqn:schec}) were obtained by fitting a Schechter luminosity function to mock galaxies and found to be $M^*_{i'-\mathrm{band}}=-20.69 \pm 0.24$ and $\alpha = -1.114 \pm 0.028$. We optimise 3D-MF to find clusters with masses $>10^{14}\Msun$ by setting the cutoff radius, $r_{co}$, to be 1Mpc based on the $r_{200}$ radius of a $3 \times 10^{14}\Msun$ cluster.  This is sufficiently large to enclose non-symmetrically shaped clusters and thus allow them to be detected, but still small enough to maximally prevent neighbouring or on-the-verge-of-merging galaxy clusters from becoming blended into one cluster detection.  P96 explored a range of cutoff radii and found the significance of their cluster detections dropped by up to 40\% when increasing $r_{co}$ beyond 1Mpc, but didn't test decreasing this window size significantly within 1Mpc. However, P96 notes that when cluster dimensions are significantly larger then the cutoff radius, the signal will obviously be truncated, though the degree depends on cluster shape.  In the Millennium Simulation analysis that follows we find our choice of $r_{co}$ is very successful at detecting clusters of masses $>1 \times 10^{14}\Msun$ and has a very useful by-product of also detecting clusters with masses as low as $\sim 10^{13}\Msun$.  As the set cutoff radius for these low mass clusters is significantly larger than their extent, it will be difficult to interpret our results for these objects as we discuss further in Section \ref{selectfcnnoerr}.  Future implementations of 3D-MF will experiment with optimising low mass detections using a varying cutoff radius.

\subsubsection{Single Galaxy Cluster Detection Criteria}\label{singlecrit}
Following 3D-MF's search of the Millennium Simulation mock lightcone catalogues, the output detected galaxy cluster list was compared with the known Millennium Simulation galaxy clusters (see Section \ref{MSgalcls}).  To match a galaxy cluster detection with known galaxy clusters, a cut was first made in a projected two-dimensional radius, set at 0.044 degrees.  This value was chosen due to the fact that 0.044 degrees at a mid-redshift range, $z \sim 0.55$, corresponds to $\sim$1Mpc, the assumed cluster cutoff radius ($r_{co}$) for an individual cluster as per Table \ref{3dmfparams}, and notably at the higher redshift end, $z=0.95$, 0.044 degrees corresponds to $\sim$1.25Mpc remaining a reasonable choice there as well.  Note that sizing this tolerance per redshift did not significantly change results.  Therefore, any galaxy cluster detections within 0.044 degrees of a known Millennium Simulation cluster were considered possible {\emph{candidates}}.  From the list of {\emph{candidates}}, each 3D-MF found cluster was matched in 3D space (RA, Dec, z) to the closest known cluster and that known cluster is thus considered detected.  The redshift component of this matching was performed within ranges of $z_{\mathrm{3D-MF}} \pm 0.1$ for redshifts $> 0.6$ (where $z_{\mathrm{3D-MF}}$ denotes the 3D-MF derived cluster redshift, and $\pm 0.1$ indicates that this matching was done within a redshift slice), and $z_{\mathrm{3D-MF}} \pm 0.1 \pm 10\% \mathrm{(of \ slice \ width)}$ for $z \le 0.6$.  Note that lower redshift clusters will be spread out in redshift bins more appreciably than higher redshift clusters in relation to the bin volume; we wanted to avoid missing proper matches between 3D-MF clusters and known mock clusters seemingly spread out due to this effect and thus a slight widening of the matching redshift parameter for lower redshift clusters was chosen.

\subsubsection{Multiple Galaxy Cluster Detection Criteria}\label{multdet}
As per the discussion in Section \ref{MSgalcls}, it can easily become complicated to define a galaxy {\emph{cluster}}.  For example, in some cases it can become difficult to realistically determine whether a multiply detected galaxy cluster is either a) one larger known galaxy cluster detected multiple times (for instance consider a merging system in a slightly 'dumbbell' shape that still has two clear lobes from the individual galaxy clusters, each lobe separate and large enough to be detected individually despite the chosen $r_{co}$ value), or b) from one detection of a cluster and a second detection is instead from smaller known nearby clusters in the lower mass ranges.  For this reason, we chose to track multiple detections in three-dimensional space, and apply grouping criteria.  Grouping 3D-MF detections to 3D-MF detections by a physically motivated linking length, and calling these linkages {\emph{multiple detections}}, we can track multiple detections within each redshift slice, as well as between redshift slices.  We wanted to avoid falsely assigning detected clusters to the wrong known galaxy clusters and to track multiply detected clusters at all mass ranges separately from false detections.  Resultantly, any 3D-MF detections within a predetermined linking length of each other were assigned a similar grouping identification number.  This linking length was chosen to be 0.75Mpc (in two dimensions: RA and Dec) based on the physically motivated assumption that if one 3D-MF detection could be exactly at the correct known centre position, and noting that clusters typically span 1.5Mpc in diameter, this would put a detection at 0.75Mpc away from the correct centre as a maximal radius by which to associate detections.  The third dimension of linking length, redshift, was selected to be $z_{\mathrm{3D-MF}} \pm 0.1$ (i.e. joint membership in an original slice and an overlapping shifted slice) and was not chosen to be 0.75Mpc as in the two-dimensional linking, simply because the resolution in redshift space is unfortunately not this high.  \cite{whitekochanek} implement a similar method and drop cluster candidates when a more likely cluster detection exists within a projected separation of $1h^{-1}\mathrm{Mpc}$ and redshift difference of $\pm 0.05$.  3D-MF's multiple detection grouping criteria are summarised in Table \ref{3dmfparams}.  It is important to note that 3D-MF detections were associated to each other and the known cluster centres were not used in this step, because we wanted to develop a method that would transfer to real data where centres are always unknown.\\
\\
The {\emph{best}} match in a multiple detection grouping is chosen to be the one that is closest to a known simulation cluster centre.  We point out to the reader that this choice is a difficult one, as there are known issues regarding centroiding; how one decides the true centre of a cluster is an interesting science topic many papers devote themselves to in their entirety.  We will provide evidence in Section \ref{centroidsec} that serves to support our choice of best match, and plan to investigate multiple matches and centroiding in future work.  Our selection function plots are free of multiple detections not considered best matches as we believe these are duplicate detections of single clusters. 

\subsubsection{False Detection Criteria}
False detections were those 3D-MF cluster detections that did not match to any known galaxy clusters after the above matching and grouping criteria were applied.  As described, the cluster finding process is repeated with original and shifted slices.  In order to ensure each false detection was not a 'faint' detection of a real structure (for example, to determine whether a false detection was actually found more significantly in a different redshift slice, perhaps due to a cutting of the cluster), and to merge original and shifted slices, the false detections from the original slices were cross referenced in RA, Dec, and redshift with real cluster detections from overlapping shifted slices (and vice versa).  The remaining false detections will be quantified and are likely due to noise fluctuations or subsisting projection effects.

\subsection{Modelling Photometric Redshift Errors}\label{zmodels}
\begin{figure}
  \includegraphics[width=8cm]{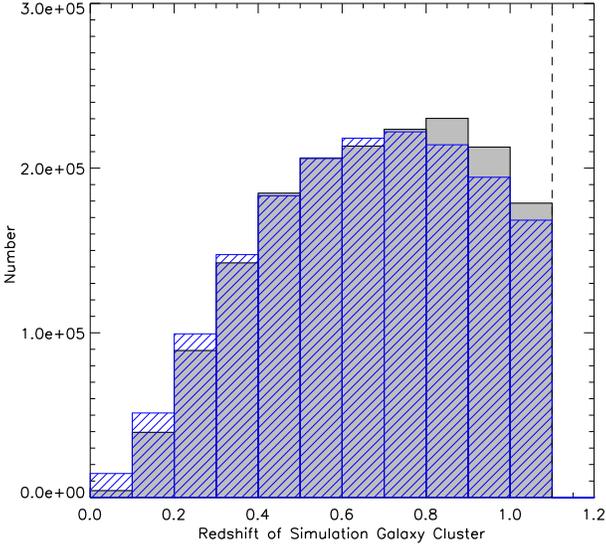}
  \caption{Redshift distribution of the 6 Millennium Simulation mock lightcones (KW07) as exact redshifts (solid histograms) and with redshift errors modelled from the photometric redshift errors of \protect \cite{cars2} (blue, dashed histograms).  Errors were derived from comparisons between the CFHTLS with VVDS and DEEP2 overlap regions; an explanation of the error modelling can be found in the text (Section \ref{zmodels}).  The data have been cut at a limiting $i'-$band magnitude of 25.5 to mimic real, observationally limited data, and are only shown out to a redshift of 1.1 as explained in the text.}
      \label{zdistnMS:magcut}
\end{figure}

Photometric redshift errors were modelled for this work by examining the area of overlap between the Canada-France-Hawaii Telescope Legacy Survey (CFHTLS) Wide field data and the VIMOS VLT Deep Survey (VVDS; \cite{vvdsspectra}) and the DEEP2 spectroscopic survey (\cite{deep2spectra}). Photometric redshifts were estimated for the CFHTLS Wide data from images provided by the CFHTLS-Archive-Research Survey (\cite{cars1}) with the method described in \cite{cars2}, and those photometric redshifts were then compared to secure spectroscopic redshifts. This process yields an accuracy of the photometric redshifts for galaxies with magnitudes $i'<24$ of $\sigma_{\Delta z/(1+z)}\approx 0.047$ after rejection of 2.8\% of outliers.\\
\\
We confine the redshift range of galaxies to $z \le 1.1$ as that is the region where our photometric redshifts are most accurate and not badly affected by biases. Common objects in the overlap regions of both CFHTLS and VVDS and CFHTLS and DEEP2 were binned according to magnitude and redshift and the photometric redshift error (i.e. the rms of $(z_{\mathrm{photometric}}-z_{\mathrm{spectroscopic}})$ after rejection of outliers) for each bin was calculated and weighted by the bin number counts.  A three-dimensional surface (magnitude, photometric redshift, and photometric redshift rms) was then found to be best fit with a 2 degree polynomial: the rms error of any new magnitude and redshift can then be extrapolated from this model (including an extrapolation of errors for magnitudes such as $i'>24$ where deeper spectroscopic redshift catalogues do not currently exist to allow more accurate modelling of this high magnitude region). Using the Millennium Simulation mock lightcone data, a galaxy's magnitude and exact redshift was used in conjunction with the aforementioned rms model to calculate a typical observational photometric redshift error rms for that particular galaxy, from which a Gaussian distribution was created. The assumption of a Gaussian error distribution is supported by the comparisons between photometric redshifts and spectroscopic redshifts mentioned above (no significant secondary peak or skewness was found as confirmed by the low number of outliers stated above and an overall vanishing photo-z bias respectively).  Randomly assigning a photometric redshift error to a given galaxy was then accomplished by randomly drawing an error value from the Gaussian distribution with the rms modelled for that galaxy.  The redshift distribution of mock lightcone galaxies before and after the redshift errors were applied are shown in Figure \ref{zdistnMS:magcut} (where the data have also been cut at a limiting $i'-$band magnitude of 25.5 to mimic real observational data, and are only shown out to a redshift of 1.1 although, as mentioned, the lightcones themselves are much deeper).  The redshift error scales with redshift; selecting redshift slices of 0.2 in width ensures our redshift slicing is larger than the rms of $\sim 0.1$ at a redshift of 1.1.

\subsubsection{Fine-Tuning 3D-MF Parameters}\label{finefine}
\begin{figure}
\includegraphics[width=8cm]{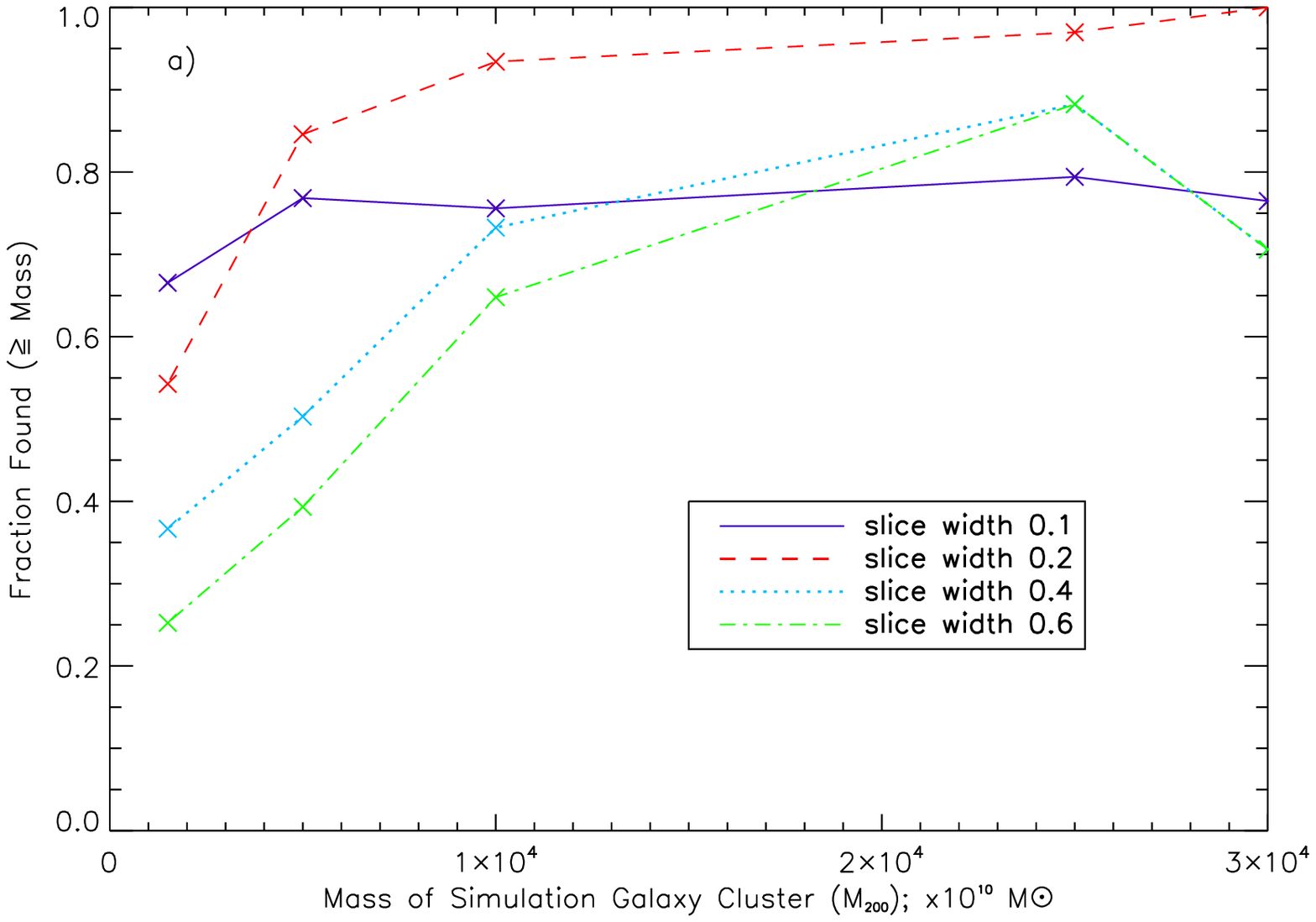} \\
\\
\includegraphics[width=8cm]{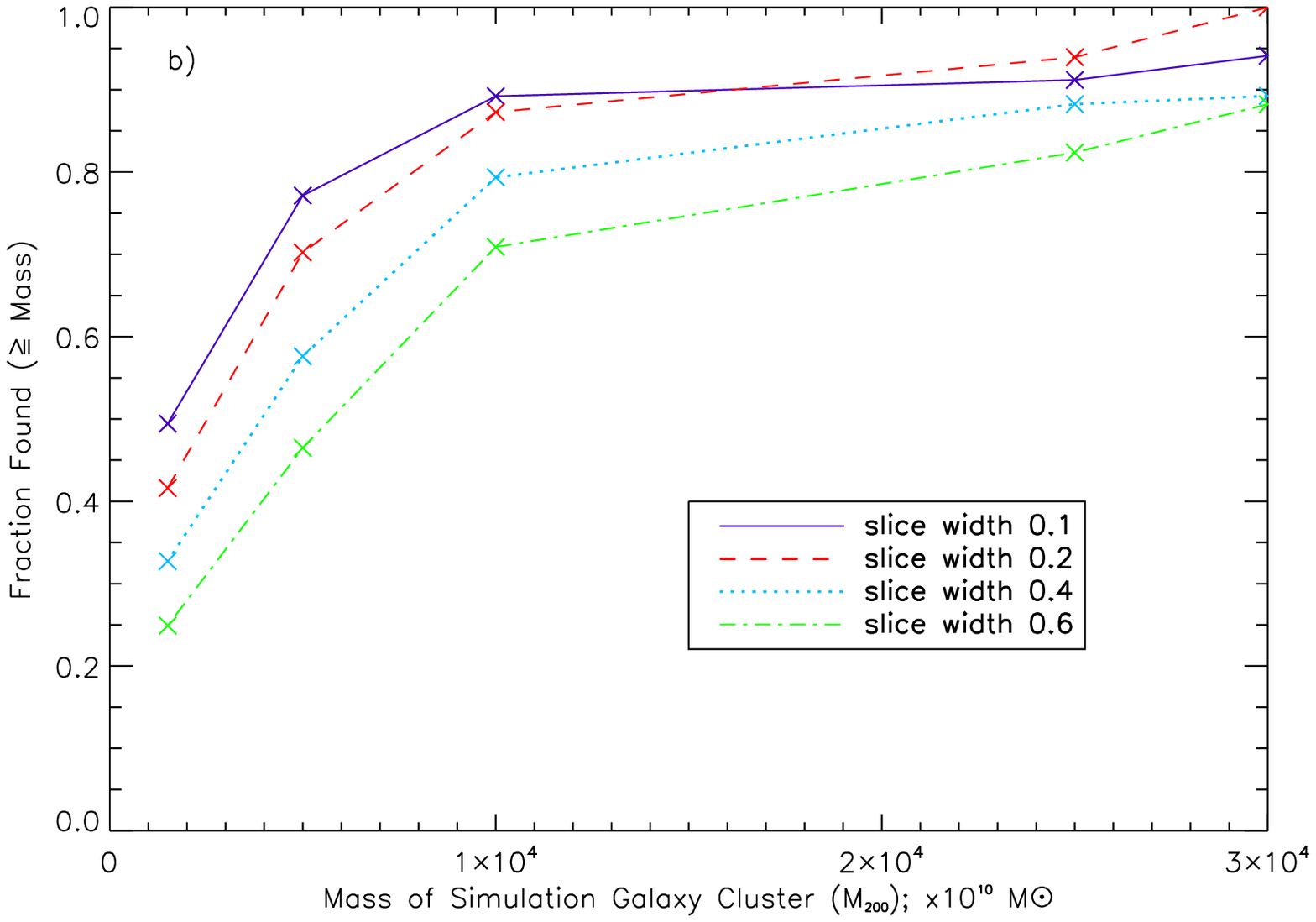} \\ 
\caption{Completeness with respect to mass, measured as a fraction of known clusters found by 3D-MF. The input data was sliced into overlapping redshift slices as explained in the text.  Overlapping slices are in each case shifted by half of the redshift slice width.  Figure \protect\ref{completenessNOerrcompletenessWerr}a presents exact redshifts while Figure \protect\ref{completenessNOerrcompletenessWerr}b includes redshift errors in the data.}
\label{completenessNOerrcompletenessWerr}
\end{figure}

Investigating the use of various redshift slice widths with 3D-MF, as well as examining the effects of introducing redshift errors into the data, and the corresponding recovery rates of galaxy clusters, leads us to be able to fine-tune 3D-MF's parameters and select an optimal redshift slicing.\\
\\
An examination of the completeness (the fraction of known clusters recovered by 3D-MF) as a function of mass {\emph{and}} redshift slice width is essential to recover as many clusters as possible.  Figure \ref{completenessNOerrcompletenessWerr}a presents a few variations of redshift slice width as evidence that an analysis of the mock lightcones with 3D-MF set to use redshift slice bin widths of 0.2 in size recovers the most known clusters with $M_{200} \ge 5\times 10^{13}\Msun$.  Repeating this analysis with photometric redshift errors shows that completeness tends to go up in almost all cases when running 3D-MF separately with various redshift slice widths (see Figure \ref{completenessNOerrcompletenessWerr}b) but the significance of detections against the background goes down (as will be shown).  \\

\begin{figure}
\includegraphics[width=8cm]{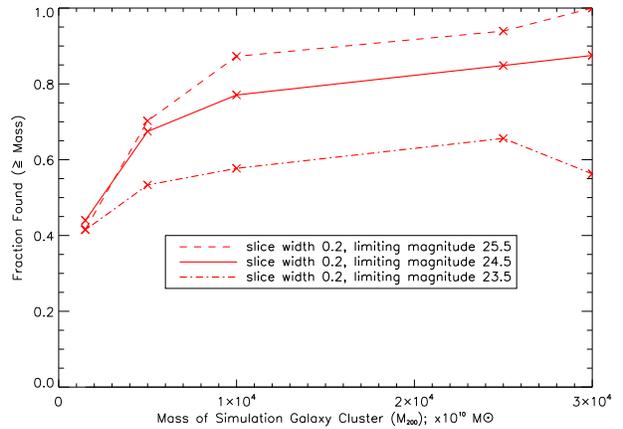} \\ 
\caption{Similar to Figure \protect\ref{completenessNOerrcompletenessWerr}b (completeness with respect to mass, measured as a fraction of known clusters found by 3D-MF), but here the width of the redshift slices is held fixed and the effect (on the fraction of galaxy clusters found) caused by changing the $i'-$band limiting magnitude of the survey is shown.}
\label{masscompletenessWerr}
\end{figure}

\begin{figure}
  \includegraphics[width=8cm]{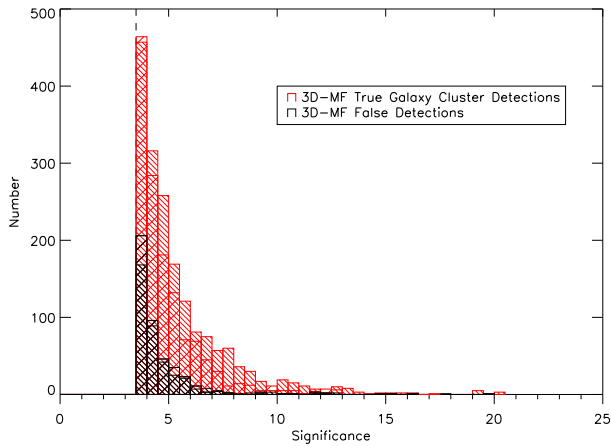} 
  \caption[]{Comparing 3D-MF's real, true galaxy cluster detections with false detections as a function of detection significance (down-sloping lines represent no redshift errors, while up-sloping lines represent data with redshift errors).}
  \label{sigmaoplotsigmaoplotERR}
\end{figure}

\noindent Maintaining a constant redshift slice width, it is of further interest to note the effect on completeness when the limiting magnitude of the survey is changed.  Figure \ref{masscompletenessWerr} presents a constant redshift slice width of 0.2 with varying survey depths of 23.5 to 25.5 $i'-$band magnitudes.  As expected, the ability to find all clusters is significantly reduced as the survey depth decreases: fewer galaxies are present in shallower data making their parent clusters improbable or impossible to optically detect.  Interpretations associated with cluster masses below $\sim 5\times 10^{13}\Msun$ will be addressed throughout Section \ref{selectfcnnoerr} and the complications in this region will be addressed.
\\
\\
Returning to a discussion of significance levels, ignoring any detections below the 3.5$\sigma$ level is a sensible decision: the probability of a 3$\sigma$ positive curvature peak from a background Gaussian field, due to chance, is 4\% (see \cite{waerbekenoise}).  Effectively, desiring a contamination due to random fluctuations of less than 1\%, we have to be more conservative and accept only those cluster detections above 3.5$\sigma$.  Figure \ref{sigmaoplotsigmaoplotERR} investigates significance and clearly shows false detections were not found at a high level of significance when compared to known clusters accurately detected by 3D-MF.  Low false detection significances are encouragingly still seen in Figure \ref{sigmaoplotsigmaoplotERR} with the introduction of photometric redshift errors (and Figure \ref{sigmaoplotsigmaoplotERR} shows, as expected, redshift errors reduce detection significances in general).  If we cut at a higher significance level, such as $6.5\sigma$, we could remove $> 95\%$ of our false detections, but we would also lose many lower significance, real detections; a trade-off must be reached and thus $3.5\sigma$ was chosen. 

\subsection{Final Selection Functions}\label{selectfcnnoerr}
\begin{figure}
  \includegraphics[width=8cm]{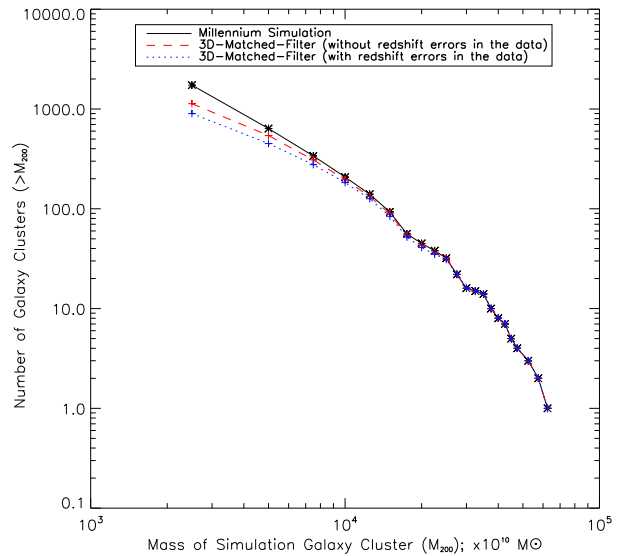} 
  \caption[]{Cumulative mass function for galaxy clusters from the Millennium Simulation KW07 catalogue (as per Table \ref{MS_CLdef}; redshift errors described in the text).}
  \label{cumfcnNOerrcumfcnWerr}
\end{figure}

\begin{figure}
  \includegraphics[width=8cm]{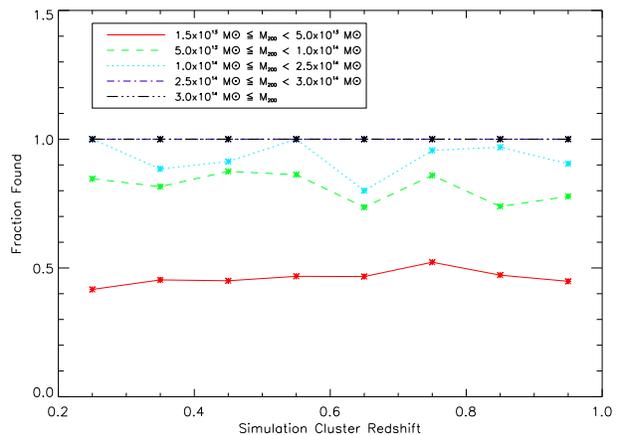} 
  \caption[]{Completeness with respect to redshift, measured as a fraction of known clusters found.}
  \label{completenessZNOerr}
\end{figure}

Having determined appropriate parameter settings, we now present 3D-MF's selection functions.  The cumulative mass function of galaxy clusters recovered by 3D-MF is presented in Figure \ref{cumfcnNOerrcumfcnWerr}.  An examination of this figure shows the recovery rate of known clusters to be very accurate above an $M_{200}$ of $\sim 5\times 10^{13}\Msun$; results that are encouraging for 3D-MF, and furthermore are consistent for all redshifts (see Figure \ref{completenessZNOerr}).  \\

\begin{figure}
  \includegraphics[width=7.8cm]{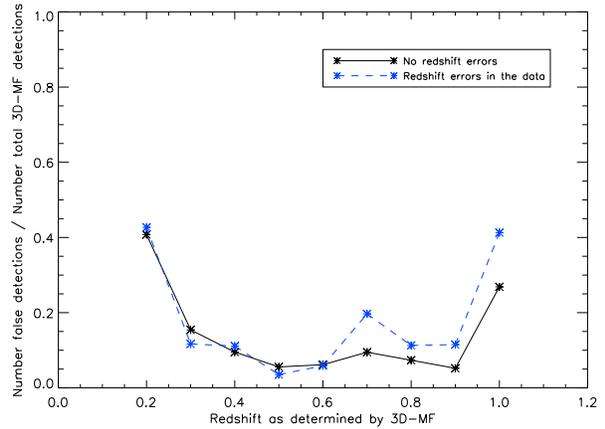}
  \caption{3D-MF false detections, as a fraction of total 3D-MF detections, shown as a function of redshift.}
  \label{FDngals}
\end{figure}

\noindent Taking the optimal redshift bin width of 0.2 (as per Section \ref{finefine}) and examining completeness with respect to redshift results in fractional recovery rates of known clusters per redshift that are constant in each mass range, with excellent recovery (i.e. $> 88\%$) for clusters with $M_{200} \ge 1.0\times 10^{14}\Msun$.  Completeness rates per mass, multiple detection rates, and false detection information are summarised in Table \ref{FDnoerrFDwitherr}. \ 3D-MF has a 100\% recovery for clusters with $M_{200} \ge 3.0\times 10^{14}\Msun$, 97\% of clusters are found with masses above $2.5\times 10^{14}\Msun$, 88\% recovery rates are seen for $M_{200} \ge 1.0\times 10^{14}\Msun$ as mentioned, and 72\% of clusters are found in even lower mass ranges (numbers quoted pertain to the presence of photometric redshift errors).  \\
\\
For clusters with redshifts in the range $0.2 \le z \le 1.0$ and $M_{200} \ge 1.0\times 10^{14}\Msun$, the galaxy cluster number density in the Millennium Simulation KW07 catalogues is $\sim 18$ galaxy clusters per square degree; 3D-MF finds $\sim 17$ galaxy clusters per square degree (using exact redshifts) and $\sim 16$ galaxy clusters per square degree (using redshifts with errors) for this same redshift and mass range.  Furthermore, for clusters with $1.5\times 10^{13}\Msun \le M_{200} < 1.0\times 10^{14}\Msun$, the galaxy cluster number density in the Millennium Simulation KW07 catalogues is $\sim 258$ galaxy clusters per square degree; using exact redshifts, 3D-MF finds $\sim 134$ galaxy clusters per square degree, and when photometric redshift errors are added, 3D-MF finds a cluster number density of $\sim 100$ galaxy clusters per square degree. \\
\\
The multiple detection rate, measured as {\emph{the fraction of total 3D-MF detections matched to known clusters that are already matched to a 3D-MF detection}}, for clusters with $M_{200} \ge 1.5\times 10^{13}\Msun$, is 36.7\% when exact redshifts are used and 36.2\% when redshift errors are added to the data. This multiple detection rate is what would be expected considering our redshift slices overlap by 0.1 in redshift: many clusters are included wholly in both original and shifted slices and, depending on their simulation redshift, are found by either preferentially matching our fiducial cluster (recall Section \ref{oldfoundations}) in the original or shifted slices, or both.\\
\\
False detection rates are reported as {\emph{the fraction of total 3D-MF detections not matched to known clusters above $1.5\times 10^{13}\Msun$}}; when simulations have photometric redshift errors added, the false detection rate increases a mere 4.4\% from the non-error photometric redshift case to 15.6\% false detections overall.  Since photometric redshift errors scatter galaxies to random redshift slices they are not expected to induce increased clustering.  False detections can be further analysed; they are shown in Figure \ref{FDngals} to be a fairly uniform function of redshift, with the exception of the redshifts at the edges of our redshift range ($z=0.2$ and 1.0).  It is likely that some of these edge redshift false detections are in fact true detections of simulation clusters with redshifts $<0.2$ or $>1.0$.  This needs to be investigated further, but note that regardless of the redshift range chosen, there will always be potential contamination from galaxy members of simulation clusters whose centres lie outside the range of interest but whose presence is enough to cause a 3D-MF detection.  Recall that we had reasons to select this redshift range: introducing redshift errors into the simulations was confined to a redshift region where photometric redshift errors could be accurately modelled (see Section \ref{zmodels}).\\
\begin{table*}
   \caption{Completeness (percent of total known clusters found by 3D-MF) and multiple and false detection rates for 3D-MF on KW07 simulation galaxy clusters (photometric redshift errors described in the text).  The Detected Galaxy Clusters section reports the number of clusters detected at least one time, the Multiple Detections section reports additional detections, and the False Detections section presents the fraction of total 3D-MF detections not matched to known clusters above $1.5\times 10^{13}\Msun$. \protect \label{FDnoerrFDwitherr}}
   \begin{tabular*}{0.8\textwidth}{@{\extracolsep{\fill}}lrrrrr}
      \hline
      \\[-1ex]
      \multicolumn{6}{l}{\bf Detected Galaxy Clusters} \\ [1ex] \hline
      \\[-1ex]
      Mass Range & Number of & \multicolumn{2}{r}{No photometric redshift errors}  & \multicolumn{2}{r}{With photometric redshift errors}  \\ 

	         & Known Clusters & Number & Completeness & Number & Completeness \\ [1ex] \hline
$\ge 3.0\times 10^{14}\Msun$ & 16   & 16   & 100.0\% &  16 & 100.0\% \\ 
$\ge 2.5\times 10^{14}\Msun$ & 32   & 32   & 100.0\% &   31 & 96.9\% \\ 
$\ge 1.0\times 10^{14}\Msun$ & 208  & 197  & 94.7\% &  184 & 88.5\% \\ 
$\ge 5.0\times 10^{13}\Msun$ & 637  & 548  & 86.0\% &  456 & 71.6\% \\ 
$\ge 1.5\times 10^{13}\Msun$ & 3240 & 1775 & 54.8\% & 1358 & 41.9\% \\ [1ex] \hline
      \\[-1ex]
      \multicolumn{6}{l}{\bf Additional Multiple Detections} \\ [1ex] \hline
      \\[-1ex]
      & & & Percent of total & & Percent of total \\ 
      & & & Cluster Detections & & Cluster Detections \\ [1ex]  \hline
      $\ge 3.0\times 10^{14}\Msun$ & &   14 &  0.4\% &   21 & \multicolumn{1}{r}{0.7\%} \\ 
      $\ge 2.5\times 10^{14}\Msun$ & &   28 &  0.8\% &   39 & \multicolumn{1}{r}{1.4\%} \\ 
      $\ge 1.0\times 10^{14}\Msun$ & &  208 &  6.1\% &  227 & \multicolumn{1}{r}{8.1\%} \\ 
      $\ge 5.0\times 10^{13}\Msun$ & &  504 & 14.8\% &  439 & \multicolumn{1}{r}{15.6\%} \\ 
      $\ge 1.5\times 10^{13}\Msun$ & & 1253 & 36.7\% & 1020 & \multicolumn{1}{r}{36.2\%} \\ [1ex] \hline
      \\[-1ex]
      \multicolumn{6}{l}{\bf False Detections} \\ [1ex] \hline
      \\[-1ex]
      $\ge 1.5\times 10^{13}\Msun$ & & 383 & \multicolumn{1}{r}{11.2\%} & 438 & \multicolumn{1}{r}{15.6\%} \\ [1ex] \hline
\\
   \end{tabular*}
\end{table*} 

\noindent As previously discussed, 3D-MF measures the significance of each detection above the background galaxy distribution (Equation \ref{eqn:sigma}).  False detections as a function of the detection significance are presented in Figure \ref{sigmaoplotsigmaoplotERR} with the significance levels at which known clusters were accurately detected: many more highly significant clusters were found which were real detections, rather than false detections.  Introducing redshift errors does lower the significance of detections, but as Figure \ref{completenessNOerrcompletenessWerr} shows, completeness remains high.

\subsubsection{Does 3D-MF Have a Proxy for Mass?}\label{proxy}
\begin{figure}
  \includegraphics[width=8cm]{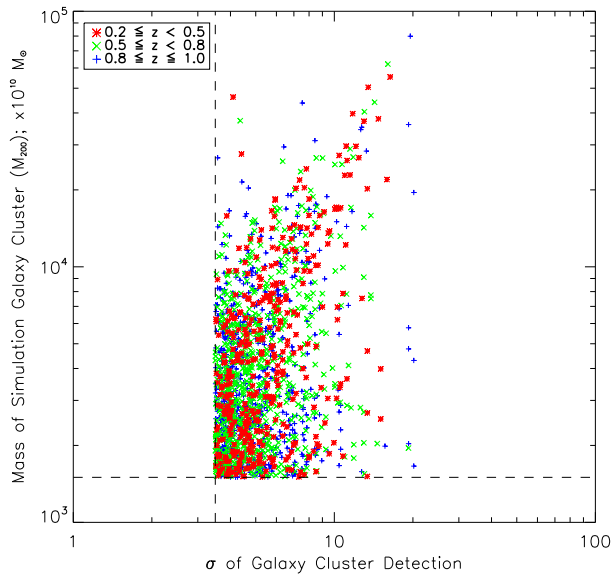}
  \caption{Significance of galaxy cluster detections in all 6 simulation lightcones.  Only detections $\ge 3.5 \sigma$ are considered actual real detections.  Known clusters are not considered below $1.5\times 10^{13}\Msun$ (a sensible mass cut, well below the divergence from expectation as shown in Figure \ref{cumfcnNOerrcumfcnWerr}).  The redshift bins in the plot do not reflect redshift slicing, but rather are found clusters binned by their redshifts.}
  \label{massvssigNOerr}
\end{figure}

\begin{figure*}
\includegraphics[width=8cm]{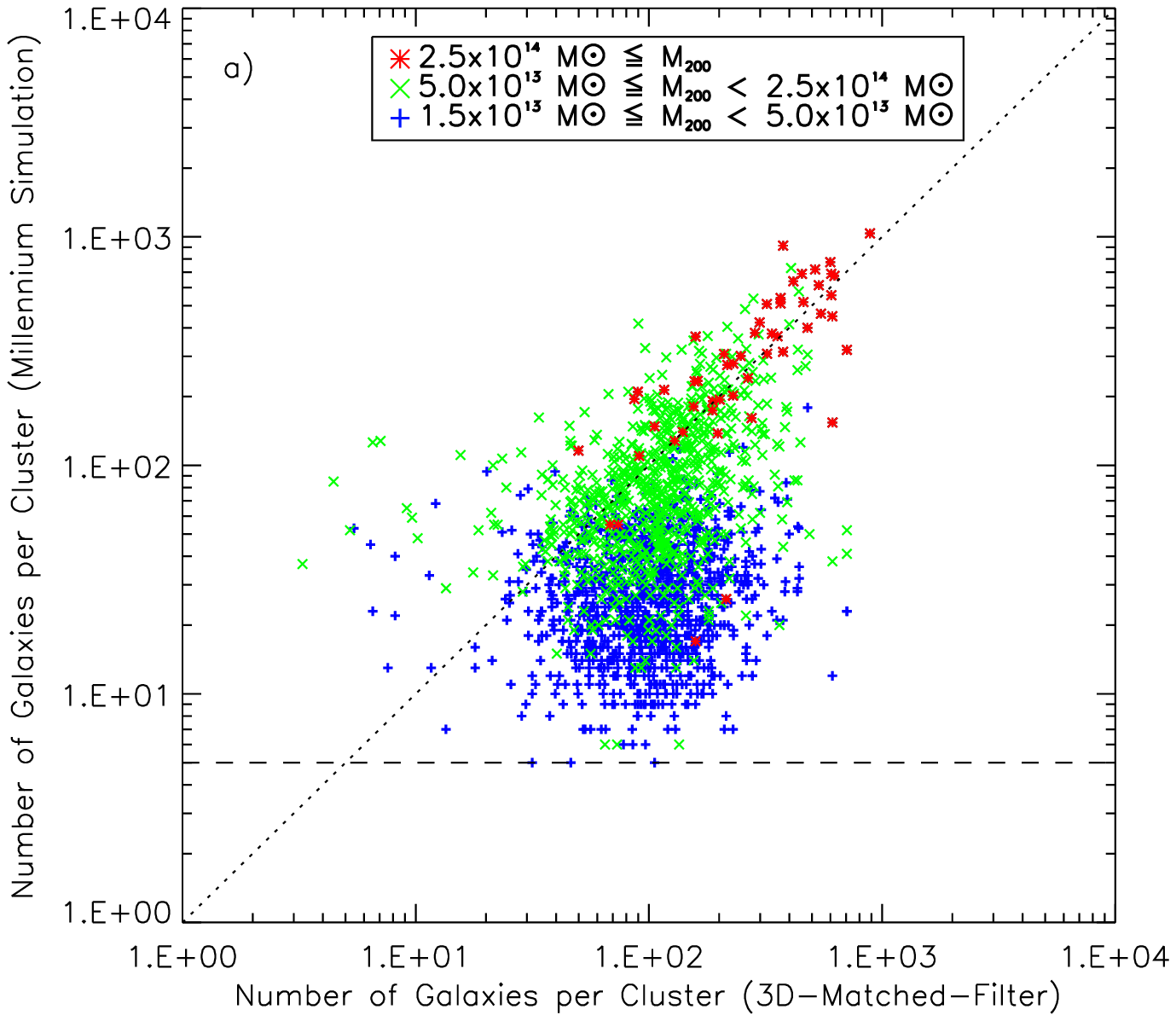} $ \ \ \ \ \ \ \ \ \ $ \includegraphics[width=8cm]{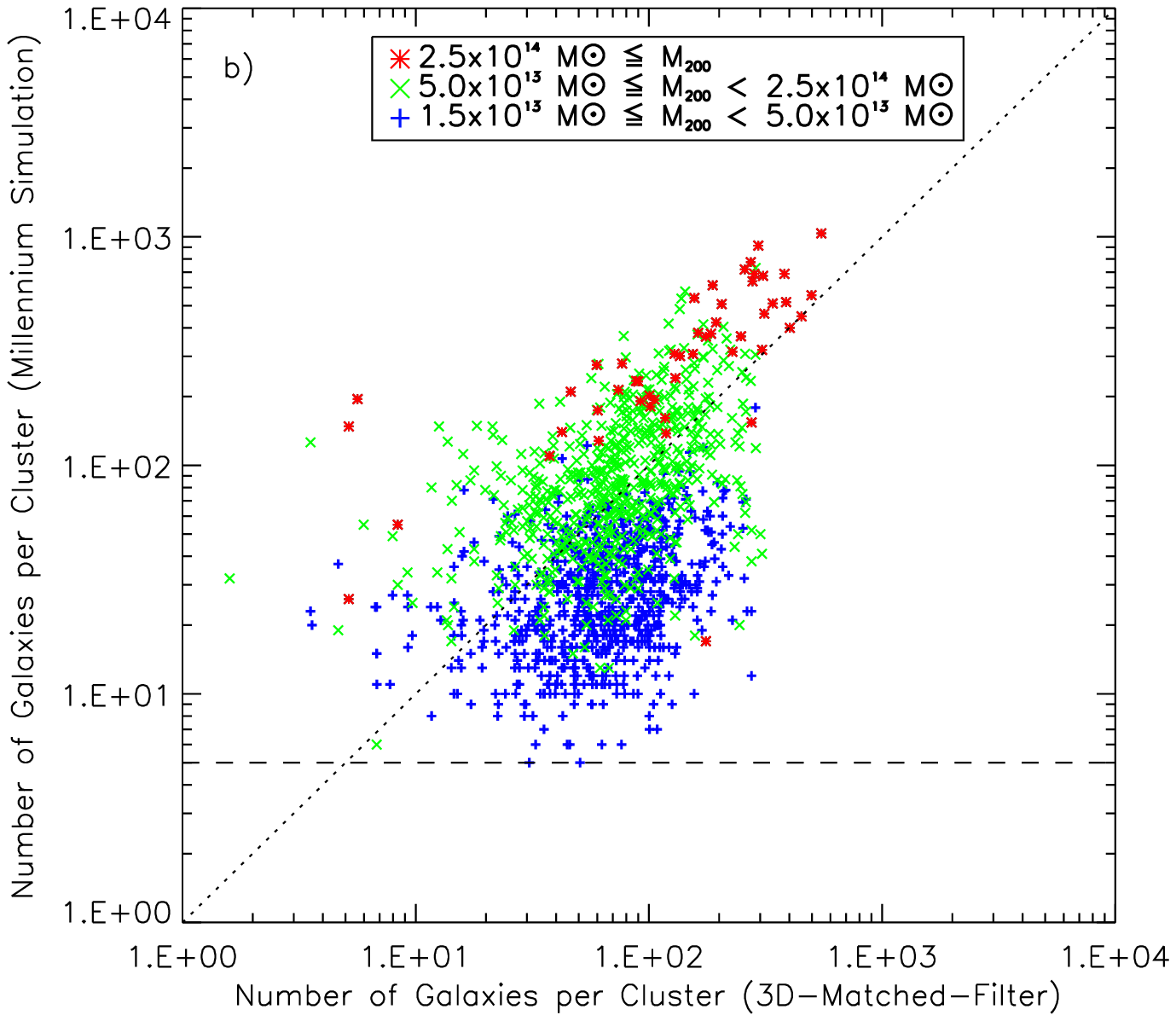} \\
\caption{The number of galaxies found per known cluster as a function of cluster mass (the dashed horizontal line denotes the definition in Table \ref{MS_CLdef} which requires clusters have $\ge 5$ galaxy members).  Figure \protect\ref{ngalsvsngalsNOerrMass}a presents exact redshifts while Figure \protect\ref{ngalsvsngalsNOerrMass}b includes redshift errors in the data.}
\label{ngalsvsngalsNOerrMass}
\end{figure*}

\begin{figure}
  \includegraphics[width=8cm]{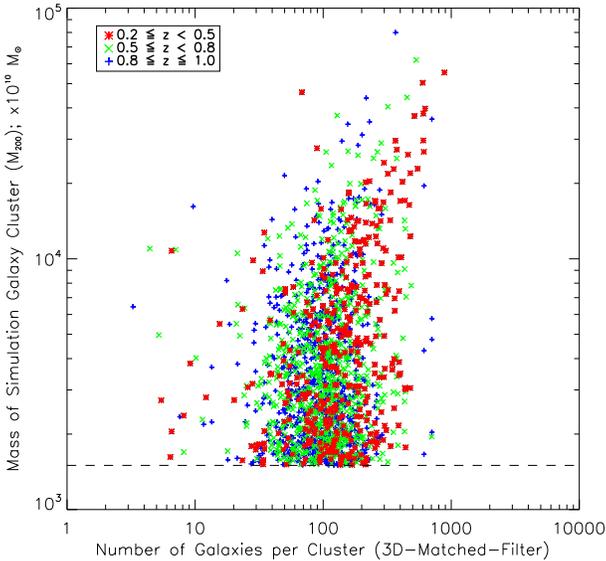} 
  \caption{The range of cluster masses and the number of galaxies associated to those cluster masses by 3D-MF.}
  \label{massvsngalNOerr}
\end{figure}

The significance of cluster detections ($\ge 3.5 \sigma$) as a function of cluster mass is presented in Figure \ref{massvssigNOerr}.  Focusing on clusters with masses $>10^{14}\Msun$ we see higher mass clusters being detected with higher significance as one might expect.
\noindent For lower mass clusters we find the detection significance is not correlated with mass and we believe this is due to a number of reasons.  We looked at the few cases of $>10 \sigma$ detections matched with known clusters of masses $<2 \times 10^{13}\Msun$ and found that while the closest known cluster to the detection was a low mass cluster, in the majority of cases galaxy interlopers from a nearby more massive, $>10^{14}\Msun$, cluster were more likely responsible for the significance of the detection.  In addition, as discussed in Section \ref{params}, 3D-MF is not fine-tuned for interpreting the detection of low mass clusters as the radial filter is currently much larger than the extent of these low mass clusters.  3D-MF assigns all galaxies within a 1Mpc radius around a peak detection to a cluster; for low mass clusters, galaxies within this radius and within a redshift slice, but outside the low mass cluster halo, will add noise to a measurement of significance.    This is demonstrated in Figure \ref{ngalsvsngalsNOerrMass} which presents the number of galaxies found by 3D-MF per known cluster as a function of cluster mass. \\
\\
The left panel of Figure \ref{ngalsvsngalsNOerrMass} (Figure \ref{ngalsvsngalsNOerrMass}a) shows the results for exact redshifts and the right panel (Figure \ref{ngalsvsngalsNOerrMass}b) shows the results when photometric redshift errors are included.  For clusters with $M_{200}>5 \times 10^{13}\Msun$, 3D-MF correctly measures the number of galaxies belonging to a cluster.  However, for lower mass clusters we see the measured galaxy cluster member counts unable to go below $\sim 30$ galaxies per cluster (blue crosses), which is the average number of galaxies that fit 3D-MF's radial profile within a 1Mpc window.  Reducing the cutoff radius, $r_{co}$, in Equation \ref{eqn:hub} would improve the association of correct galaxies to low mass clusters but this would significantly reduce the high rate of success shown for higher mass cluster detections.\\
\\
The introduction of redshift errors moves the points in Figure \ref{ngalsvsngalsNOerrMass}a leftward, meaning that as cluster members become scattered in redshift space, 3D-MF associates less of them to their parent cluster, as would be expected.  Interestingly, for the lowest mass clusters, which are the most difficult to detect, this means galaxy number counts are less likely to be overestimated while remaining detectable with 3D-MF.  \\ 
\\
Examining the number of galaxies per cluster, as determined by 3D-MF, as a potential proxy for mass is shown in Figure \ref{massvsngalNOerr}.  Again focusing on the clusters with masses $>10^{14}\Msun$ we see a weak redshift dependent trend: the most massive clusters have the most numerous members.  For the low mass clusters however, owing to the incorrect assignment of too many galaxies to those clusters, we see no correlation.  \\
\\
We will investigate not having a clear proxy for mass via 3D-MF in future work.  P96, D07 and O1-4 assume that all light after background subtraction follows the Schechter function and that this can be expressed as a luminosity equivalent to the number of $L^*$ galaxies (which they refer to as $\Lambda_{cl}$), with good results, but accurate galaxy number counts per cluster is less of an issue with the lower cluster densities and higher mass clusters their methods typically find.  We plan to further investigate our currently overestimated cluster galaxy members in the low mass ranges, by looking at galaxy number counts per cluster, and potentially optimising a variable cutoff radius in the 3D-MF search window.\\ 
\\
Methods complimentary to 3D-MF would work as mass estimators as well; weak gravitational lensing techniques for example can be utilised on real data to measure cluster masses \citep[as in][]{Hoekstra}.  Alternatively, spectroscopic redshifts could be utilised to precisely identify galaxy cluster members and potentially lead to determining a better contamination fraction in the current relationship between 3D-MF galaxy cluster member number counts and cluster mass.  X-ray information is known to be an excellent proxy for mass \citep{smith}, and Sunyaev-Zel'dovich (SZ) information could be added to tell us more about the structure of clusters along the line-of-sight (as in \cite{SZ1} and \cite{SZ2} for example).

\subsubsection{Cluster Catalogue Purity}\label{centroidsec}
\begin{figure}
  \includegraphics[width=8cm]{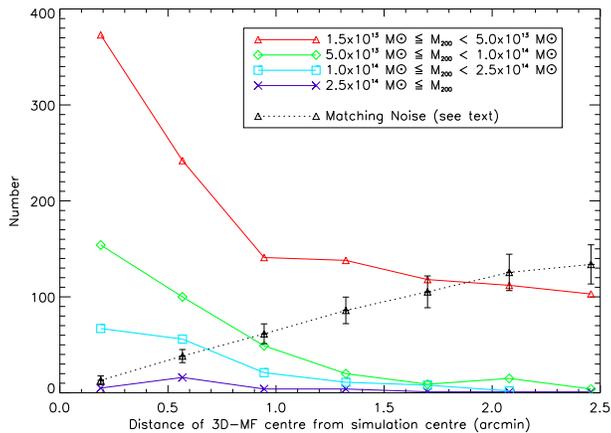}
  \caption{Distance of 3D-MF derived cluster centres from predetermined, known simulation centres (as defined in Section \ref{MSgalcls}). The dotted line shows results from 3D-MF derived clusters randomised in RA and Dec before matching to known simulation clusters according to our matching criteria: over 80\% of the contribution here is from the lowest (i.e. $1.5\times 10^{13}\Msun \le M_{200}<5.0\times 10^{13}\Msun$) mass range (see text for further discussion).}
  \label{centroidplot}
\end{figure}

\noindent As previously discussed, it is not an easy task to define the centre position of a cluster in an automated way.  By construction, 3D-MF does not have a cluster centre predefined.  Other cluster finding methods, such as those mentioned in Section \ref{intro}, use the position of the brightest cluster galaxy member \citep{koester}, or the centre of the distribution of red galaxies for example.  We choose the centre of a galaxy cluster to be based on the luminosity and radial distribution of cluster galaxies.  As shown in Figure \ref{ngalsvsngalsNOerrMass}, for clusters with $M_{200}>5 \times 10^{13}\Msun$, 3D-MF correctly associates galaxy members to their parent cluster, and we thereby expect the cluster centroid to be determined by this method with reasonable accuracy.  For lower mass clusters however, owing to interloping galaxy members from nearby clusters (interlopers that fall within a radial profile centred near the lower mass cluster), the centroid that we measure is rather difficult to interpret.  \\
\\
Recall that a 3D-MF cluster detection within 0.75Mpc of another simulation matched 3D-MF cluster detection (a 3D-MF detection that was chosen to be the match as it was closer to the known cluster centre) is considered a secondary detection.  This further complicates the issue of centroiding, as it is not clear whether the 3D-MF centroid for clusters with secondary detections should be calculated from the galaxies associated to both the primary and secondary detection or just the primary, especially considering the fact that clusters often have substructures (and perhaps in reality don't have a clear centre).  In the analysis that follows we choose the primary detection as the centroid.  \\
\\
Figure \ref{centroidplot} shows the number of 3D-MF derived clusters matched to known simulation clusters as a function of the distance between their centres.  For all mass ranges we find there are fewer 3D-MF cluster centre positions with increased radius from known cluster centres, showing that a 3D-MF centroid measurement is frequently quite accurate.  The elevated plateau at a matching radius $>1$ arcminute, seen in the $1.5\times 10^{13}\Msun \le M_{200}<5.0\times 10^{13}\Msun$ mass range, indicates some level of contamination in our low mass matched sample.  \\

\noindent To investigate this, we randomised the cluster centroids found by 3D-MF in RA and Dec and repeated our matching to known simulation cluster centre positions.  The average random matches for all cluster mass ranges, from 50 randomised trials, is presented in Figure \ref{centroidplot} as the dotted line, which we call matching noise.  Over 80\% of the contribution to this matching noise comes from low mass clusters ($1.5\times 10^{13}\Msun \le M_{200}<5.0\times 10^{13}\Msun$) and is thus a major source of the contamination in the low mass sample's plateau (red, triangle line) in Figure \ref{centroidplot}.  This is what we would expect: low mass (and thus small) clusters contribute to the matching noise increasingly at larger radii from known KW07 centres since randomising their 3D-MF found positions should uniformly place their detections across the field, and with any size of matching window there will always be some of these low mass clusters (of high number density) matched to a known cluster. \\
\\
The matching noise we have shown is an overestimate of the real matches due to noise because randomising the 3D-MF cluster positions in RA and Dec eliminates properly identified multiple detections of single clusters, thereby overestimating single detections and maximally increasing the matching due to chance incorrect matches.  For example, a cluster that is detected by 3D-MF in one original redshift slice and also in the overlapping redshift slice would be considered a multiply detected cluster.  After randomisation of the cluster positions, this would now appear as two separate cluster detections in the original and shifted redshift slices.  To correct for this, we use our multiple detection rates of 36.2\% and conservatively report a maximal upper limit on our false detections of 24.3\% (i.e. [(Total number of false detections + $(1-0.362) \ \times$ Total number counts due to an upper limit of matching noise)/Total 3D-MF detections]).  As discussed at the beginning of this section, we believe the false detection rate (especially for clusters of masses $\ge 5.0\times 10^{13}\Msun$) is closer to 15.6\%, but for the higher number densities of lower mass clusters this could be up to 24.3\%.  \\
\\
Due to the increased sensitivity of 3D-MF over two-dimensional matched-filter methods, we believe 3D-MF is correctly recognising the significance of low mass clusters above background noise, but due to excess galaxy number counts per cluster, and oversized fiducial cluster filters, has trouble correctly choosing a cluster centroid for these low mass clusters.  We can safely confirm from this centroid analysis that we have chosen a physically motivated and sensible matching radius within which to properly match 3D-MF's detections with known simulation clusters. Interestingly, the two-dimensional matched-filter of \cite{whitekochanek} successfully implements an iterative centroiding technique: a logical next step for improving the current implementation of 3D-MF and its ability to determine cluster centres. 

\section{Canada-France-Hawaii Telescope Legacy Survey Deep Field Cluster Catalogues Generated by 3D-MF}\label{3DMFonCFHTLS}
\subsection{Canada-France-Hawaii Telescope Legacy Survey Deep Data and Catalogue Generation}
For the current work we use the four $1 \times 1$ square degree Canada-France-Hawaii Telescope Legacy Survey (CFHTLS) Deep fields (D1,2,3,4) and COSMOS data which were presented within the CFHTLS-Archive-Research Survey; we refer the reader to \cite{cars1} and \cite{cars2} for details on the data processing and multi-colour catalogue creation. \\
\\
Our five-band multi-colour and photometric redshift catalogues are based on CFHTLS $i'-$band object detections. The seeing for this band varies between 0.71 and 0.82 arcseconds within the four fields and we reach a $1\sigma$ limiting magnitude of $i'_{\mathrm{AB,lim}}\approx 29.5$ within a diameter of twice the seeing disk. Photometric redshifts were estimated with the publicly available code \texttt{BPZ} (see \cite{benitez2000} and the prescription in \cite{cars2}). In the fields D1, D2 and D3 we quantified the accuracy of our photometric redshifts with spectroscopic redshifts from the VVDS (\cite{vvdsspectra}), $z$COSMOS (\cite{lilly2007}), and DEEP2 (\cite{deep2spectra}) respectively.  Within the magnitude range of $17< $i'-$band <24$ we estimate unbiased photometric redshifts, and after rejecting problematic sources (we cut with the \texttt{BPZ} \texttt{ODDS} parameter for sources with \texttt{ODDS}$>0.9$; see also \cite{cars2}), we find a photometric - spectroscopic redshift scatter of $\sigma \approx 0.033$ of the quantity $\Delta z=(z_{\mathrm{phot}}-z_{\mathrm{spec}})/(1+z_{\mathrm{spec}})$. Our outlier rate with $|\Delta z| > 0.15$ is $1.6\%$.  Our final Deep catalogues contain $\sim1.23\times10^{5}$ galaxies per square degree with $i'-$band magnitudes $<25.5$ (compared to the simulations with $z<1.2$ and the same magnitude limit, which had $\sim1.59\times10^{5}$ galaxies per square degree).  

\subsection{CFHTLS Deep Galaxy Cluster Catalogues Courtesy of 3D-MF}\label{clustercat}
3D-MF was set up according to the parameters in Table \ref{3dmfparams} but an appropriate $M^*$ and $\alpha$ were derived from the CFHTLS Deep fields and found to be $M^*_{i'-\mathrm{band}}= 22.46 \pm 0.15$ and $\alpha = -1.005 \pm 0.021$.  3D-MF was then run on the CFHTLS Deep fields.  Table \ref{clustercatlist} contains 8 random entries (2 per field) from our CFHTLS Deep galaxy cluster catalogue; the entire catalogue is available upon request to the authors.  From our Millennium Simulation tests we would expect to see $\sim 16\%-24\%$ false positives in this catalogue, distributed mostly in the lower mass ranges according to the selection functions in Section \ref{3DMFonMSnozerr}.  Using our multiple detection criteria, we found 37.6\% of Deep detections were duplicate detections of clusters (comparable to the $\sim 36\%$ multiple detection rate found from our Millennium Simulation tests).  Grouping Identification Numbers in Table \ref{clustercatlist} are numbered flags: cluster detections within 0.75Mpc of each other (recall Section \ref{multdet} for details) in the same redshift slice are flagged with identical {\emph{Same Redshift}} numbers, and cluster detections within a projected 0.75Mpc of each other in overlapping redshift slices are separately flagged with identical {\emph{Overlapping Redshift}} grouping identification numbers (which can be propagated through more than one overlapping redshift slice if still within 0.75Mpc in projected radii of each other in overlapping slices).  Note that Grouping Identification Numbers in both the {\emph{Same Redshift}} and {\emph{Overlapping Redshift}} numbering restart at zero for each Deep field. \\
\\
We use the significance of our detections to select the best galaxy cluster candidate from among multiple detections, and excise the remaining multiple detections from the following discussion and analysis.  The redshift distribution of 3D-MF found Deep galaxy clusters is shown in Figure \ref{deepcompALL}.  A comparison to other published CFHTLS Deep cluster lists via older matched-filter methodology ensues.
\begin{table*}
   \caption{3D-MF Derived CFHTLS Deep Galaxy Cluster Catalogue. \protect \label{clustercatlist}}
   \begin{tabular*}{1.0\textwidth}{@{\extracolsep{\fill}}ccccccc}
      \hline
      \\ [-2ex] 
      \multicolumn{7}{l}{\bf CFHTLS Deep Galaxy Clusters} \\ [1ex] \hline
      \\ [-2ex] 
      Deep  & RA & Dec & Redshift & Significance & \multicolumn{2}{c}{Grouping Identification Number} \\ 
      Field &    &     &    &  ($\sigma$) & Same Redshift & Overlapping Redshift \\ [1ex] \hline
      \\ [-2ex] 
      D1 & 02:25:07.536 & -04:01:47.892 & 0.2 & 7.16 & - & - \\
      D1 & 02:24:26.040 & -04:52:29.676 & 0.9 & 5.21 & 2 & 58 \\
      D2 & 09:58:54.720 &  01:54:57.744 & 0.4 & 5.35 & - & - \\
      D2 & 09:58:53.759 &  02:14:16.188 & 0.5 & 5.15 & - & 31 \\ 
      D3 & 14:17:22.802 &  52:54:52.920 & 0.2 & 7.27 & - & - \\ 
      D3 & 14:18:19.440 &  53:05:58.200 & 0.5 & 5.61 & - & - \\
      D4 & 22:14:39.120 & -18:09:41.760 & 0.6 & 5.23 & - & 37 \\ 
      D4 & 22:16:08.400 & -18:11:48.840 & 0.4 & 6.73 & - & - \\ [1ex] \hline
      \\ [-2ex]
      \multicolumn{7}{c}{...complete catalogue available upon request...} \\ [1ex] \hline
      \\ 
   \end{tabular*}
\end{table*}
 
\subsection{Comparison to Other Published CFHTLS Deep Clusters Found Using Matched-Filter Methods}\label{compare3dMFO07}
\begin{figure}
  \includegraphics[width=7cm]{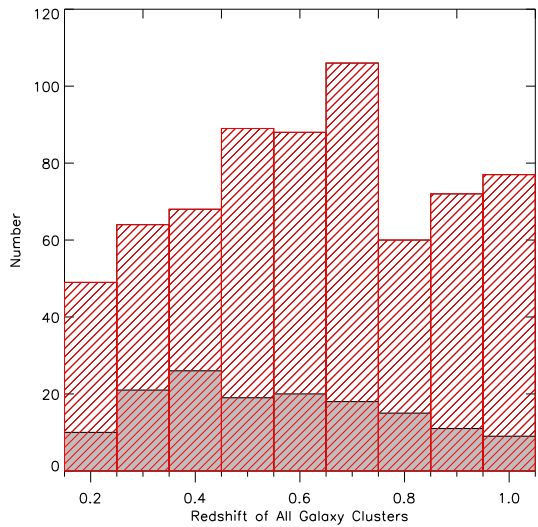}
  \caption{Redshift distribution of all 3D-MF found (dashed up-sloping histogram) and \protect \cite{olsen1stcfhtls} found (shaded histogram) galaxy clusters in CFHTLS Deep fields 1 through 4.}
  \label{deepcompALL}
\end{figure}

\cite{olsen1stcfhtls} (herein O3) published a matched-filter derived cluster list of the CFHTLS Deep fields.  A subsequent paper in 2009 (\cite{grove2ndcfhtls}; O4) examined running their matched-filter method (which does not utilise photometric redshifts) separately on different wavelength bands and then merging lists with the 2007 paper, a technique similar to P96's efforts to reduce false detection rates. Since the resulting lists are not too different, we choose to compare the $i'-$band 2007 O3 cluster list to a cluster list found by 3D-MF, using the same $i'-$band data, in this paper.  We focus on the $0.2 \le z \le 1.0$ range for clusters, keeping in line with our acceptable photometric redshift region for galaxies and our slicing methodology.  The redshift distributions of the O3 cluster list, after their $3.5\sigma$ cut, for all four CFHTLS Deep fields is shown overlayed on the 3D-MF results in Figure \ref{deepcompALL}.  Drastically fewer galaxy clusters were found by O3 at all redshifts; 3D-MF finds $\sim 170$ galaxy clusters per square degree (well within reason compared to the Millennium Simulation analysis cluster number densities in Section \ref{selectfcnnoerr}) compared to $\sim 40$ galaxy clusters per square degree in O3. \\

\begin{figure}
  \includegraphics[width=7cm]{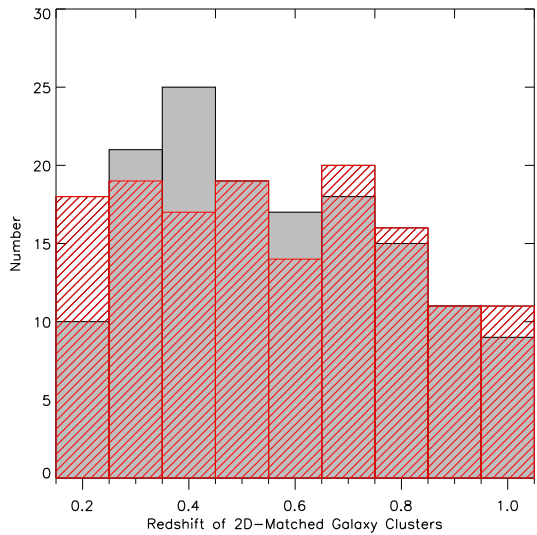}
  \caption{The redshift distribution of only those galaxy clusters found using {\emph{both}} 3D-MF and O3 methods, matched to each other in two-dimensional space (see text for details).  3D-MF results are represented by the dashed up-sloping histogram, and O3 results are the shaded histogram.}
  \label{deepcomp2D}
\end{figure}

\begin{figure}
  \includegraphics[width=7cm]{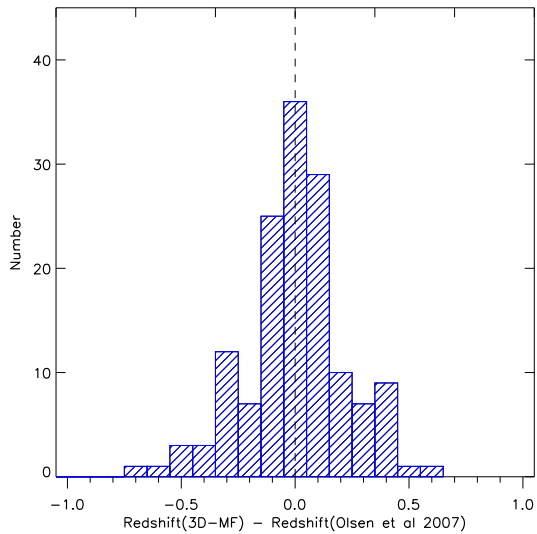}
  \caption{Differences in assigned galaxy cluster redshifts for clusters found using both 3D-MF and O3 methods (using two-dimensional matching as described in Section \ref{compare3dMFO07})}
  \label{deepcompZdiff}
\end{figure}

\noindent In order to match the clusters found using both methods, a two-dimensional tolerance radius of 0.044 degrees (akin to the matching strategy in Section \ref{singlecrit}) was placed around each O3 cluster centre position and the closest 3D-MF detection in RA and Dec was considered a match.  Figure \ref{deepcomp2D} shows the redshift distribution, and Figure \ref{deepcompZdiff} presents the redshift differences, of the two-dimensional matches between the two cluster lists.  \\

\begin{figure}
  \includegraphics[width=7cm]{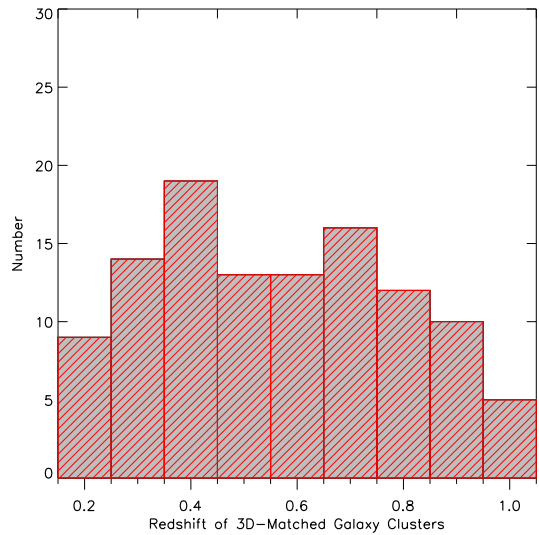}
  \caption{Redshift distribution of galaxy clusters found using both 3D-MF and O3 methods, matched to each other in {\emph{three}}-dimensional space (see text for details).  The vertical scale of Figure \protect \ref{deepcomp2D} is duplicated here for comparison purposes.}
  \label{deepcomp3D}
\end{figure}

\noindent As mentioned in Section \ref{assnz}, older matched-filter methods tend to wrongly estimate the redshift of clusters.  O3 searches the entire input catalogue (i.e. no redshift slicing) for a match to a fiducial cluster sized to match what would be expected at a particular redshift, and then repeats the process with a slightly re-sized fiducial cluster size (to match what would be expected for a cluster that existed at a slightly different redshift); O3 requires each cluster to be detected in two neighbouring fiducial cluster re-sizing searches in order to be considered further.  This affects higher redshifts disproportionately, as they are less likely to be found in equally-stepped fiducial cluster-resizing bins than lower redshift clusters; higher redshift space of equivalent bin widths covers much more volume.  Figure \ref{deepcomp2D} shows the likely result of this: the O3 cluster sample is skewed toward lower redshifts. \\\\
A second separate matching was performed between the 3D-MF Deep cluster list and O3; this time a {\emph{three}}-dimensional tolerance range was considered around each O3 cluster.  There were less matches, as seen in Figure \ref{deepcomp3D}, again skewed toward the lower redshift end of the distribution.  \\
\begin{table}
      \caption{A Comparison of CFHTLS Deep galaxy clusters found with 3D-MF and published CFHTLS Deep galaxy clusters from O3 \protect \citep{olsen1stcfhtls}. Cluster redshift ranges of $0.2 \le z \le 1.0$ are considered, keeping in line with this work, and galaxy clusters present in both catalogues with similar redshifts (as explained in the text) are reported again in the 'O3 \& 3D-MF' column. \protect \label{deepcomptable}}
      \begin{tabular*}{0.45\textwidth}{@{\extracolsep{\fill}}cccc}
         \hline
	 \\ [-2ex]
	 \multicolumn{4}{l}{\bf 3D Cluster Comparison} \\ [1ex] \hline
	 \\ [-2ex]
	 CFHTLS     & 3D-MF & O3 & O3 \\
	 Deep Field &       &         & {\bf \&} 3D-MF \\ [1ex] \hline
	 \\ [-2ex]
	 D1    &   161 & 43      & 32 \\ 
         D2    &   162 & 38      & 31 \\ 
	 D3    &   190 & 38      & 25 \\ 
	 D4    &   160 & 30      & 23 \\ [1ex] \hline
	 \\ [-2ex]
	 Total D1-4 & 673 & 149 & 111 \\ \hline
	 \\ [-2ex]
\\ 
      \end{tabular*}
\end{table} 

\noindent Comparing both the 3D-MF and O3 CFHTLS Deep cluster lists, there were 145 clusters found by both 3D-MF and O3 out of 149 total O3 clusters (97\%) with the two-dimensional matching criteria described.  528 additional galaxy clusters were found with 3D-MF. As mentioned, the matching process between 3D-MF clusters and O3 clusters was repeated considering a further third dimension in redshift space (see Table \ref{deepcomptable}); in this case there were 111 clusters found by both 3D-MF and O3, suggesting 34 of the O3 clusters (23\%) were not assigned the correct redshift by their algorithm. \\
\\
The completeness has been shown to be vastly different between the two methods: 3D-MF is 100\% complete down to $\sim 3\times 10^{14}\Msun$, $\sim$90\% complete down to $1\times 10^{14}\Msun$ and the false detections are likely concentrated even further down the cluster mass function at the low mass end.  It was discussed in general in Section \ref{oldhighfalse} that older matched-filter methods have false detection rates of $\sim 30-65\%$, suffer from line-of-sight projection issues, wrongly estimate redshifts, and have not been tested on Millennium Simulation-like, realistic, extensive simulations.

\section{Conclusions}
3D-Matched-Filter (3D-MF) is an optimised, automated galaxy cluster finder, ideal for large optical astronomical datasets with multi-wavelength band coverage.  It utilises a fiducial galaxy cluster's radial and luminosity profiles and searches the data for matches to this over a wide range of redshifts.  Future surveys, such as LSST and JDEM, can exploit 3D-MF's automated methodology and statistical approach to produce complete and reliable galaxy cluster catalogues.  3D-MF is an improvement over other matched-filter techniques due to several improvements, including:

\begin{itemize}
\item the cutting of input data into overlapping redshift slices to examine it piecewise, significantly reducing line-of-sight projections; 
\item an implementation of mask-handling capabilities, improving edge-effects, and cluster detections near bright stars or saturated pixels;  
\item an accurate modelling of data-dependent background galaxy counts; 
\item the development of a new peak detection pipeline;
\end{itemize}

as well as a list of other parameters that can be fine-tuned according to the dataset being examined.  \\
\\
The Millennium Simulation mock catalogue lightcones were used to extensively test and improve 3D-MF, and selection functions for the algorithm were presented.  Redshift errors mimicking real data were modelled and added to the simulations and their effect on the selection functions was derived.  With redshift errors, and focusing on the cluster range $0.2 \le z \le 1.0$, 3D-MF was found to recover 100\% of known galaxy clusters with an $M_{200} \ge 3.0\times 10^{14}\Msun$; 97\% of clusters with an $M_{200} \ge 2.5\times 10^{14}\Msun$; 88\% of clusters with an $M_{200} \ge 1.0\times 10^{14}\Msun$; and 72\% of clusters with an $M_{200} \ge 5.0\times 10^{13}\Msun$.  36\% of detections were multiple detections of clusters.  False detections were determined to be occuring at a rate of 15.6\% of the total cluster detections, and a subsequent analysis showed this to be concentrated in the low significance, low galaxy number ($\lesssim 50$) per cluster, and likely lower mass ($M_{200} < 1\times 10^{14}\Msun$) range (potentially increased by noise up to a conservatively reported rate of 24\%). After selection functions were quantified (and the effects of adding redshift errors to the catalogues were analysed), 3D-MF was run on the four CFHTLS Deep fields.  \ 3D-MF finds $\sim 170$ galaxy clusters per square degree in the Deep dataset: over 400\% more, with a much lower false detection rate, and higher accuracy of redshift determination for true clusters, than found by other authors using two-dimensional matched-filter methods on the same $i'-$band data.  \\
\\
For future work, there are subtle adjustments to 3D-MF that we are examining.  A non-passively evolving {\emph{k}}-correction could be applied, taking into consideration the effect of variations in galaxy types.  More interestingly perhaps, as shown in \cite{popesso}, galaxy clusters are often better fit with two Schechter functions as opposed to one: implementing this result may further improve 3D-MF.  The radial profile used in 3D-MF's radial filter could also be tuned to try to find more low mass, smaller clusters (if possible).  Different filter bands could be used in 3D-MF when analysing the redshift slices, and results compared with the $i'-$band used herein, or spectroscopic redshifts could be utilised to confirm cluster detections and precisely identify galaxy cluster members.  It would be sensible to try to extend the high redshift range of the detection capabilities of 3D-MF as well, although this method is ultimately limited by the build up of the luminosity function at redshifts $\gtrsim 1.2$.  \\
\\
The Millennium Simulation has proved invaluable in refining and understanding 3D-MF as an automated optical cluster finder.  A few points were raised that we continue to study in fastidious detail: finding a 3D-MF derived proxy for cluster mass, refining the number of galaxies associated to a cluster, and improving cluster centroid determinations are examples.  In the future we will introduce a weak gravitational lensing analysis to elucidate cluster masses and we are also investigating multi-wavelength coverage in X-ray and SZ data to refine cluster information.  

\section*{Acknowledgements}
We thank the reviewer for useful comments and feedback.  MM also thanks Gabriella De Lucia for helpful discussions on the Millennium Simulation.  This work involves observations obtained with MegaPrime/MegaCam, a joint project of CFHT and CEA/DAPNIA, at the Canada-France-Hawaii Telescope (CFHT) which is operated by the National Research Council (NRC) of Canada, the Institut National des Sciences de l'Univers of the Centre National de la Recherche Scientifique (CNRS) of France, and the University of Hawaii. This work is based in part on data products produced at TERAPIX and the Canadian Astronomy Data Centre as part of the Canada-France-Hawaii Telescope Legacy Survey, a collaborative project of NRC and CNRS.  HH was supported by the European DUEL RTN, project MRTN-CT-2006-036133.  TE is supported by the German Ministry for Science and Education (BMBF) through DESY under the project 05AV5PDA/3 and the German Science Foundation (DFG) through the project TR33 'The Dark Universe'.  The Millennium Simulation databases used in this paper and the web application providing online access to them were constructed as part of the activities of the German Astrophysical Virtual Observatory.  

\small
\bibliographystyle{mn2e}
\balance

{\small{\bsp}}
\end{document}